\documentstyle[aps,epsf,epsfig,prb]{revtex}

\newlength{\bxwidth}\bxwidth=2.5 truein

\newlength{\fight}
\newcommand\ltdash{\raise-1.8pt\hbox{$\scriptscriptstyle |$}}
\newcommand \be  {\begin{equation}}
\newcommand \ee  {\end{equation}}
\newcommand \bea {\begin{eqnarray} }
\newcommand \eea {\end{eqnarray}}

\begin{document}
\title{Atomic model of susy Hubbard operators}
\author{J. Hopkinson and P. Coleman}
\address{Center for Materials Theory,
Department of Physics and Astronomy, 
Rutgers University, Piscataway, NJ 08854, USA.
}
\date{\today}
\maketitle
\begin{abstract}
We apply the recently proposed susy Hubbard operators [P. Coleman, C. P{\'e}pin and J. Hopkinson, Phys. Rev. B {\bf{63}}, 140411(R) (2001)] to an atomic model.  In the limiting case of free spins, we derive exact results for the entropy which are compared with a mean field + gaussian corrections description.  We show how these results can be extended to the case of charge fluctuations and calculate exact results for the partition function, free energy and heat capacity of an atomic model for some simple examples.  Wavefunctions of possible states are listed.  We compare the accuracy of large N expansions of the susy spin operators [P. Coleman, C. P{\'e}pin and A. M. Tsvelik, Phys. Rev. B {\bf{62}}, 3852 (2000); Nucl. Phys. B {\bf{586}}, 641 (2000)] with those obtained using `Schwinger bosons' and `Abrikosov pseudo-fermions'.  For the atomic model, we compare results of slave boson, slave fermion and susy Hubbard operator approximations in the physically interesting but uncontrolled limiting case of N$\rightarrow$2.  For a mixed representation of spins we estimate the accuracy of large N expansions of the atomic model.  In the single box limit, we find that the lowest energy susy saddle-point reduces to simply either slave bosons or slave fermions, while for higher boxes this is not the case.  The highest energy saddle-point solution has the interesting feature that it admits a small region of a mixed representation, which bears a superficial resemblance to that seen experimentally close to an antiferromagnetic quantum critical point.  
\end{abstract}

{\centering{
{1. INTRODUCTION\\}}}

\vskip1pc

One of the interesting challenges arising in systems exhibiting strong electronic correlations is the ability to adequately develop a microscopic description of metallic physics in regions where Fermi liquid theory breaks down.  The observations of $\frac{B}{T}$ scaling{\cite{trovarelli}}, linear resistivity
over three decades{\cite{custers}} in temperature coinciding with a
T$\ln[\frac{T_0}{T}]$ divergence in the specific heat capacity{\cite{trovarelli}} in
YbRh$_2$Si$_2$ is reminiscent of properties shown by CeCu$_{6-x}$Au$_x$ which additionally exhibits $\frac{E}{T}$ scaling{\cite{schroeder}}.  In these heavy fermion systems close to an antiferromagnetic quantum critical point only one energy or length scale seems to be important--which  generally only occurs in theories which are
below their upper critical dimension.{\cite{why}}  Evidence for a qualitatively similar phase diagram for the high-T$_c$ cuprates has been advanced by Tallon and Loram{\cite{tallon}} who have assembled a large amount of experimental data on T* and interpret it in terms of an underlying quantum critical point, presumably responsible for the linear T dependence seen in the normal state of these systems that has given rise to the marginal Fermi liquid phenomenology of Varma{\cite{Varma}} and others.  Recent $\mu$SR measurements by Panagopoulos et al{\cite{panag}} and an as yet unsubstantiated report of high temperature superconductivity in a FET of CaCuO$_2$ by Schon et al{\cite{schoen1}} have supported this interpretation.
Again, the culprit for this unusual metallic behaviour is believed to be an underlying quantum critical point.  If the
entire Fermi surface fundamentally breaks down in this region, one
might imagine that the competition between  spins trying to
magnetically order and those trying to form heavy quasiparticles
could lead to a new kind of quasi-particle excitation. {\cite{why}}  
That is, if
the Fermi temperature approaches 0 as we approach a quantum critical
point, we may see the formation of novel kinds of states; perhaps
states of higher symmetry classes could become important even at
these low temperatures. The key question that must be addressed is: how do spin and charge interact at the brink of
magnetism?

Theoretical approaches to this problem are hindered by the
difficulty of capturing the profound transformation in  spin
correlations that develops at the boundary between antiferromagnetism 
and  paramagnetism.
Usually we 
model this behavior by representing the spin
as a boson in a magnetic phase,\cite{arovas} or as a fermion in a paramagnetic
phase,\cite{coleman84} but by making this choice, the character of 
spin and charge excitations which appear in an approximate field theory  is
restricted  and lacks the flexibility to 
describe the co-existence of 
strong magnetic correlations within a paramagnetic phase.

These considerations have
motivated the development of new methods to describe the spin and
charge excitations of a strongly correlated material which avoid 
making the choice between a bosonic or fermionic spin.\cite{gan,pepin,georges,ngai} Recently, we introduced a new representation of the Hubbard operators{\cite{us}} which attempts to treat magnetism and paramagnetism on an equal footing within an approximate field theory allowing a description of physics in terms of operators which obey canonical commutation relations, yet avoid forcing us to specify the nature of a system's ground state.  In this approach, we introduce vector fields 
\be
F_a = (f_1,...,f_N, \phi) \hspace{2 cm} B_a = (b_1,...,b_N, \chi)
\ee
where $f_{\sigma}$ and $b_{\sigma}$ are Abrikosov pseudo-fermions\cite{abrikosov} and Schwinger bosons\cite{arovas} respectively, while $\phi$ and $\chi$ are slave bosons\cite{slaveboson} and slave fermions\cite{slavefermion}.  For $|a> \hspace{3 mm} \in {|\sigma>,|0>}$, the Hubbard operators can be written\cite{us} 
\be
|a><b| = X_{ab} = B^{\dagger}_a B_b + F^{\dagger}_a F_b
\ee
To guarantee that the Hubbard operator representation is irreducible,
we need to set the values of the linear and quadratic Casimirs of the group (here $g= Diag[1\dots 1,-1]$),

\begin{equation}\label{casimirs}
C^{( 1)} = \hbox{Tr}[{ X}] = Q,\qquad 
C^{( 2)}= \hbox{Tr}[XgX] = Q(N-1-Y).
\end{equation}
The first constraint corresponds to summing the total number of spin and hole states at a site (henceforth labeled by Q), while the latter describes the degree of antisymmetry of a given representation (which is fixed by Y = h-w (see Fig.1)). In terms of canonical operators, we can express Q and Y as:
\begin{equation}\label{constraint1}
Q = n_{{b}}
+ n_{{\phi}} + n_{{f}} + n_{{\chi}},
\end{equation}
\begin{equation}\label{constraint2}
Y = n_{{\phi}} + n_{{f}} - (n_{{b}} + n_{{\chi}}) + \frac{1}{Q}[\theta, \theta^{\dagger}],
\end{equation}
where
$\theta = \sum_{\sigma}{b}_{\sigma }^{\dagger}{f}_{\sigma } - {\chi}^{\dagger}{\phi}$ is an operator interconverting fermions and bosons for the corner state.  While the physically interesting case (Fig.1 (a)) corresponds to Q=1, Y=0 and N=2, one can hope to learn new qualitative features from large N expansions about different symmetry classes as has been done previously for Q = Q$_0$, Y = 1-Q$_0$ ((Fig.1(c))slave fermions--magnetism); and Q = Q$_0$, Y = Q$_0$ - 1 ((Fig.1(d))slave bosons-paramagnetism).  One now can also treat cases like (Fig.1(b) Y=0 --paramagnetism + magnetism?).

\vskip1pc
\begin{figure}[here]
\begin{center}
\includegraphics[scale=0.5]{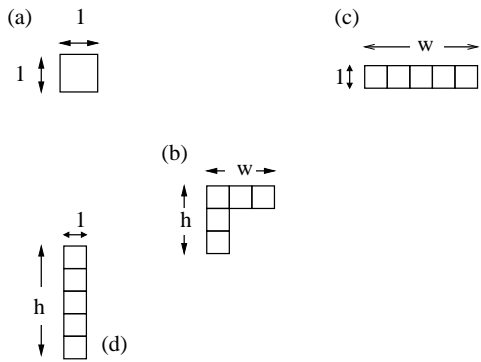}
\vskip 0.2truein
\caption{\label{figure 1}  (a) Fundamental representation $(Q,Y)= (1,0)$, (b) L-shaped Young tableau 
corresponding to the spin representation generated by  supersymmetric
Hubbard operators.  The asymmetry ${Y}=h-w $ and 
$Q$ is the number of boxes, (c) Young tableau for fully symmetric representation
corresponding to Slave fermion limit (d) Fully antisymmetric, slave boson limit. 
}
\end{center} \end{figure}

In this paper we first consider the limit of free spins with strong correlations in which charge fluctuations are totally suppressed.  In this limit, Q becomes the total number of spins ($n_b + n_f$) and the energy of the empty state is taken to infinity.  We derive exact results for the entropy of a general {Q,Y}.  Using a mean field approach plus gaussian corrections, we compute the entropy of the general free spin case and compare this with $\frac{1}{N}$ expansions of the exact results (Stirling's approximation).  We then compare the supersymmetric spin formalism{\cite{pepin2000,coleman}} with earlier large N approximations.    Representations of spins in terms of N-component `Schwinger' bosons{\cite{auerbach,arovas}} and `Abrikosov pseudo-fermions'{\cite{abrikosov}} become exact in the large N limit, but in the physical limit N $\rightarrow$ 2, provide only approximate descriptions of the physics.   Calculation of the exact entropy of free spin states, allows us to estimate the error inherent in mean field descriptions, mean field plus gaussian corrections, and self-consistent mean field + correction approximations to each of the three methods.  Surprisingly, self-consistent approximations seem to not always do as well as mean field approximations at low N.

Re-introducing charge fluctuations, we next treat an atomic model.  In such a model there is an excitation energy for a spin to leave its localized site, creating an empty or vacuum state.  The Hamiltonian for such a system is particularly simple, being given by simply the energy difference between the unoccupied site and the occupied spinful state.  In terms of Hubbard operators, this can be written:
\be
H = E_dX_{\sigma \sigma} = E_d Q - E_dX_{00} = -E_dX_{00}
\ee
where $E_d$ is the energy of the localized level, and we have dropped a constant term in the last term for convenience.  Now, we have charge fluctuations in addition to spin fluctuations allowed by the system.  For small N we can write exact wavefunctions corresponding to possible states of the system and their associated energies.  Then, we can write correspondingly exact free energies, entropies and heat capacities for such a system.  These exact values allow us at low N to compare the efficacy of the supersymmetric Hubbard operator formalism with that of previously existing slave boson{\cite{slaveboson}} and slave fermion{\cite{slavefermion}} approximations.  We find that even in this uncontrolled regime, bosonic descriptions (slave fermions) provide the best approximation to spin states, while fermionic descriptions of the spin (slave bosons) provide the best approximation to charge states at N=2.  The lowest energy saddle-point of the susy solution at N=2 corresponds to this solution, while the highest energy saddle-point generates a phase diagram which bears a superficial resemblance to the phase diagram of the cuprates and includes a small region where the solution is neither fermionic nor bosonic but mixed.  For higher N, we show how one can generate the exact partition function and estimate the accuracy in calculations of the entropy and heat capacity of this new large N approach which allows representations of the spin neither bosonic nor fermionic in character.  

\vskip1pc
{\centering
{2. FREE SPIN LIMIT}\\}

\vskip1pc
First, let us consider the free spin.  This corresponds to taking the limiting case when we send $E_d \rightarrow -\infty$.

\vskip1pc

{\centering{A. Entropy}\\}

\vskip1pc

{\centering{\it{ 1. Exact Result}}\\}

\vskip1pc

  We are left with SU(N) spins as the only states of our system.  To count the number of states available, we make use of the fact that the dimension of an irreducible representation of SU(N) can be expressed as:\cite{schonstead}
\be
\lambda_N = \prod_{i,j} \frac{(N + j - i)}{g_{ij}}
\ee
where $g_{ij}$ is the hook length of the (i,j) box of the Young tableau (see Fig. 2 a)) defined as the number of neighbors to the right and bottom of the box (see Fig. 2 b)).  For the specific case of L-shaped Young tableaux, this yields,
\bea
\lambda_N &=& \frac{N}{h+w-1}\prod_{i=2}^h(\frac{N+1-i}{h-i+1})\prod_{j=2}^w(\frac{N+j-1}{w-j+1}) \nonumber \\ &=& \frac{1}{h+w-1}\frac{N!}{(N-h)!(h-1)!}\frac{(N+w-1)!}{(w-1)!N!} \nonumber \\ &=& \left(\begin{array}{c}N \\ h \end{array}\right)\left(\begin{array}{c}N + w - 1 \\ w \end{array}\right)\frac{wh}{N(w + h - 1)}
\eea
\begin{figure}[here]
\begin{center}
\includegraphics[scale=1.0]{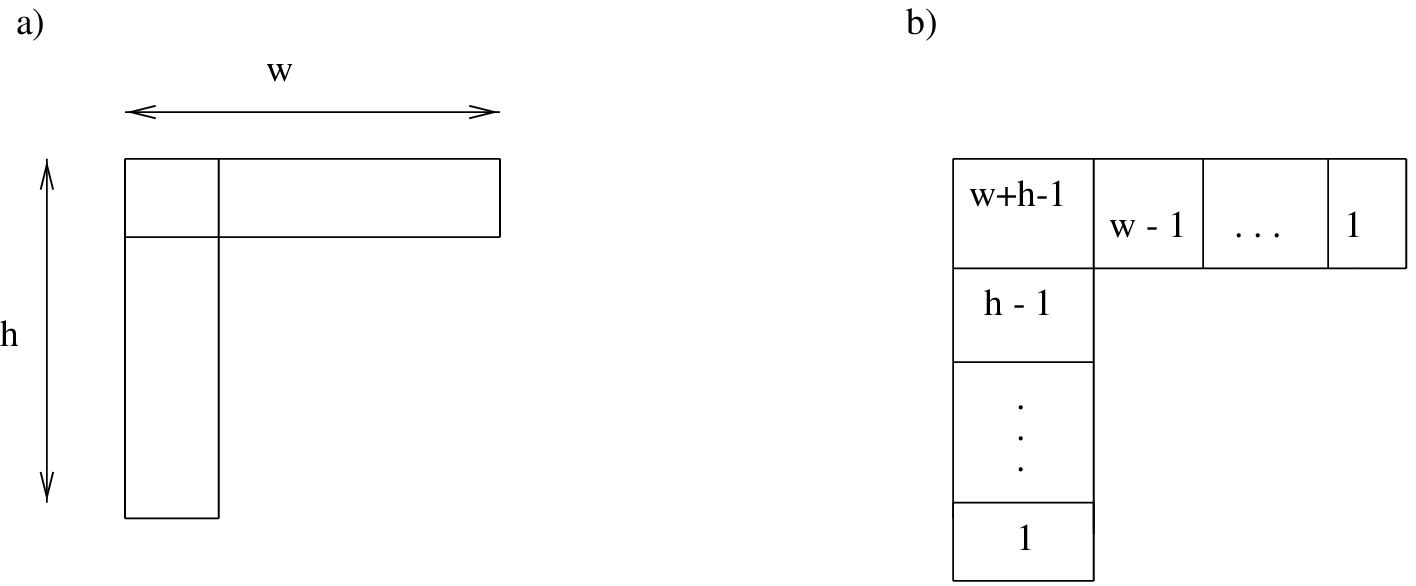}
\vskip 0.2truein
\caption{\label{figure 2}  (a) An L-shaped Young tableau with w boxes horizontally, and h boxes vertically.  (b) The hook length of each box, defined as the number of neighbors to the right and bottom of the box itself. {\cite{schonstead}} 
}
\end{center} \end{figure}

  From this formula, one can deduce an exact expression for the entropy depending only on N (SU(N)), w (width of the Young tableau), and h (height of Young tableau).  This equation is:
\be
S = \ln\left[\left(\begin{array}{c}N \\ h \end{array}\right)\left(\begin{array}{c}N + w - 1 \\ w \end{array}\right)\frac{wh}{N(w + h - 1)}\right]
\ee
Since we work in the large N limit it would be useful to know the leading dependence of this exact result.  We can use Stirling's approximation to expand the factorials, i.e.
\be
\ln(x!) = x \ln(\frac{x}{e}) + \frac{1}{2}\ln(2\pi x)
\ee
which, after some work, implies that 
\be
S_{Stirling} = N(S_f[\tilde h] + S_b[\tilde w]) - \frac{1}{2}( \ln(2\pi N \tilde w(1 + \tilde w - \frac{1}{N})) + \ln(2\pi N \tilde h (1 - \tilde h))) + \ln(\frac{w h}{N (w + h - 1)})
\ee
where $\tilde h = \frac{h}{N}, h = \frac{Q + Y + 1}{2},  \tilde w = \frac{w}{N}, w = \frac{Q - Y + 1}{2}$ and
\bea
&S_f[x] = -[x\ln x + (1-x)\ln(1-x)] \nonumber \\ &S_b[x] = -[x\ln x - (1+x)\ln(1+x - \frac{1}{N})]
\eea
In order to express h, and w as functions of N, we can remove the leading dependence of Q and Y on these variables and expand in it.  We can choose the new variable (other than N) in such a way that its sign agrees with that of Y, and its maximal amplitude is 1 for the special case when $Q = \frac{N}{2}$.  That is, we define the variable $\xi$ such that $w = \frac{N}{4} (1 - \xi) + \frac{1 + \xi}{2}$ and $h = \frac{N}{4}(1 + \xi) + \frac{1 - \xi}{2}$.  With this assignment, for the case we will consider ($Q = \frac{N}{2}$), $-1 \le \xi \le 1$ and $Y = (\frac{N}{2} -1)\xi$.  The physics at $N = 2$ will be invariant of the value of $\xi$.  It does have the problem that $N \rightarrow \infty$ and $\xi \rightarrow \pm 1$ do not commute.

Following the above-developed procedure, we express the exact entropy of the description as ($Q = \frac{N}{2}$):

\be
\frac{2S_{exact}}{N} = \frac{2}{N}\ln\left(\left(\begin{array}{c}N \\ h \end{array}\right)  \left(\begin{array}{c}N + w - 1 \\ w \end{array}\right)\frac{w h}{N (\frac{N}{2})}\right)
\ee
and expansion of this according to Stirling's approximation gives us:

\bea
\frac{2S_{Stirling}}{N} &=& -(\frac{1}{2} + \xi (\frac{1}{2} - \frac{1}{N}))\ln((\frac{1}{4} + \frac{1}{2N}) + \xi(\frac{1}{4} - \frac{1}{2N})) \nonumber \\ &-& (\frac{3}{2} - \xi(\frac{1}{2} - \frac{1}{N}))\ln((\frac{3}{4} - \frac{1}{2N}) - \xi(\frac{1}{4} - \frac{1}{2N}))  - (\frac{1}{2} + \xi(\frac{1}{N} - \frac{1}{2}))\ln((\frac{1}{4} + \frac{1}{2N}) \nonumber \\ &+& \xi(\frac{1}{2N} - \frac{1}{4})) + (\frac{5}{2} + \xi(\frac{1}{N} - \frac{1}{2}))\ln((\frac{5}{4} - \frac{1}{2N}) + \xi(\frac{1}{2N} - \frac{1}{4})) - \frac{2}{N}\ln(\pi N) -\frac{1}{N}\ln(1-\frac{1}{N}) 
\eea

extracting terms of order $\frac{1}{N}$ from inside the logs, this becomes the leading order approximation:

\bea
\frac{2S_{Stirling,lead}}{N} &=& \frac{-1}{2}(1 + \xi(1 - \frac{2}{N}))\ln(1 + \xi) - \frac{1}{2}(3 + \xi(-1 + \frac{2}{N}))\ln(3 - \xi) \nonumber \\ &-& \frac{1}{2}(1 + \xi(-1 + \frac{2}{N}))\ln(1 - \xi) + \frac{1}{2}(5 + \xi(-1 + \frac{2}{N}))\ln(5 - \xi) - \frac{2}{N} \ln(\pi N) - \frac{2}{N}
\eea
\vskip1pc

{\centering{\it{2. Mean field + Gaussian Corrections}\\}}

\vskip1pc
(a){\it{Susy.}}
Since we cannot enforce the constraints $Q = Q_0$ and $Y = Y_0$ rigorously at the mean field level, (Y has a four-fermion term of order $\frac{1}{N}$), we would like to see how inclusion of these terms consistently affects the saddle point approximation we make.  For free spins, inclusion of the constraints as Lagrange multipliers leads to an action of the form
\be
S = \int d\tau (\sum_{\sigma}\bar f_{\sigma}(\partial \tau + \lambda + \zeta)f_{\sigma} + \sum_{\sigma}\bar b_{\sigma}(\partial \tau + \lambda - \zeta)b_{\sigma} + \frac{\zeta}{Q}\sum_{\sigma,\beta}(\bar b_{\sigma} f_{\sigma} \bar f_{\beta}b_{\beta} - \bar f_{\sigma}b_{\sigma}\bar b_{\beta}f_{\beta}) - \lambda Q_0 - \zeta Y_0
\ee
The total free energy then is made up of five terms to order $\frac{1}{N}$,
\be
F_{tot} =\int {\mathcal {D}}[f,\bar f,b,\bar b]e^{-S} = F_{mft} + F_{\frac{1}{N} g. c.}= N F_f + N F_b + F_{\eta}  - \lambda Q - \zeta Y + F_{\delta \lambda_f} + F_{\delta \lambda_b}
\ee
 where the last three terms correspond to calculating contributions due to 

\be
\Pi_{\eta} = \frac{Q_0}{2\zeta} + \includegraphics[scale=1.0]{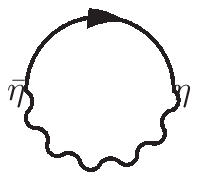}, \Pi_{\delta\lambda_f} = \includegraphics[scale=0.8]{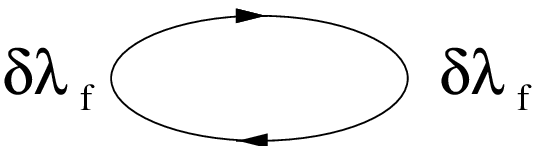},
 and \hspace{1mm}\Pi_{\delta\lambda_b} = \includegraphics[scale=1.0]{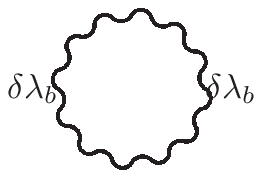},
\ee
where the two terms of order N would ordinarily define the location of the saddle point, and the three latter terms be evaluated at the location of the mean field saddle point.  As this would correspond to setting $\frac{1}{Q}<[\theta,\theta^{\dagger}]> = 0$ at the saddle point, we choose to include $F_{\eta}$ in determining the location of the saddle point itself.  This has the effect of coupling the gaussian fluctuations in $\lambda_f$ and $\lambda_b$ such that the sum $F_{\delta \lambda_f}$ + $F_{\delta \lambda_b}$ is replaced by a single term $F_{\delta_{\lambda_f},\delta_{\lambda_b}}$.  In the above, $\lambda_f = \lambda + \zeta$ and $\lambda_b = \lambda - \zeta$ arise from Lagrange multipliers $\lambda$ and $\zeta$ which have been introduced to enforce $Q = h + w - 1$ and $Y = h - w$ respectively at the mean field level.  $\eta$ is a Hubbard-Stratonovich field introduced to decouple the O($\frac{1}{N}$) term from Y.  Details of the calculation are given in Appendix A. 
Grouping all our terms from the Gaussian approximation to the entropy (calculated in Appendix A), we have
\bea
F &=& -NT\ln(1 + e^{-\beta \lambda_f}) + NT\ln(1-e^{-\beta \lambda_b})+ T \ln(\frac{\tilde n_b + \tilde n_f}{2 \tilde n_b \tilde n_f})  - \lambda (h + w) - \zeta (h - w) \nonumber \\ &+& \frac{T}{2}\ln((2 \pi)^2( N \tilde n_b (1 + \tilde n_b) N \tilde n_f (1-\tilde n_f) - n_{\alpha} (1-n_{\alpha})( N \tilde n_b (1 + \tilde n_b)+  N \tilde n_f (1-\tilde n_f) )))
\eea
Since there is no energy term here $F = -TS$ to calculate the entropy we simply need to divide by the temperature although we can also perform the constrained derivative $S = -\frac{\partial F}{\partial T}|_n = -\frac{\partial F}{\partial T} - \frac{\partial F}{\partial \lambda} \frac{\partial \lambda}{\partial T}|_n$.
\bea
\frac{2S_{mfc}}{N} &=& 2\ln(1 + e^{-\beta \lambda_f}) - 2\ln(1 - e^{-\beta \lambda_b}) + 2\frac{\beta \lambda_b}{N}w + 2\frac{\beta \lambda_f}{N}h)- \frac{2}{N}\ln(\frac{\tilde n_b + \tilde n_f}{\tilde n_b \tilde n_f}) \nonumber \\ &-& \frac{1}{N}\ln((2 \pi)^2( N \tilde n_b (1 + \tilde n_b) N \tilde n_f (1-\tilde n_f) - n_{\alpha} (1-n_{\alpha})( N \tilde n_b (1 + \tilde n_b)+  N \tilde n_f (1-\tilde n_f) )))  + \frac{2}{N}(\ln(2))
\eea
where we have used the extra $- \beta\lambda$ from the constraint on Q to create the third to last term as shown in Appendix A.  Here we have $n_{\alpha} = \frac{1}{e^{\beta 2 \zeta} + 1}$, $\tilde n_f = \frac{1}{e^{\beta \lambda_f} + 1}$ and $\tilde n_b = \frac{1}{e^{\beta \lambda_b} - 1}$ where $\lambda_f = \lambda + \zeta$ and $\lambda_b = \lambda - \zeta$ are the Lagrange multipliers as they affect fermions and bosons respectively.  The last term we remove as the factor of two from the double counting of the corner box.  Since $2\tilde h = q + y + \frac{1}{N}$ and $2\tilde w = q - y + \frac{1}{N}$, the extra $\lambda$ contribution allows us to neatly group prefactors in terms of $\tilde h$ and $\tilde w$.  That is:
\bea
\frac{2S_{mfc}}{N} &=& -2(1-\tilde h) \ln(1 - \tilde n_f) - 2\tilde h \ln \tilde n_f + 2(1 + \tilde w) \ln(1 + \tilde n_b) - 2\tilde w \ln \tilde n_b- \frac{2}{N}\ln(\frac{\tilde n_b + \tilde n_f}{\tilde n_b \tilde n_f})  \nonumber \\ &-&\frac{1}{N}\ln((2 \pi)^2( N \tilde n_b (1 + \tilde n_b) N \tilde n_f (1-\tilde n_f) - n_{\alpha} (1-n_{\alpha})( N \tilde n_b (1 + \tilde n_b)+  N \tilde n_f (1-\tilde n_f) ))) 
\eea
Taking $\frac{\partial F_{mft}}{\partial \tilde n_f} = 0 =\frac{\partial F_{mft}}{\partial \tilde n_b}$ where the mean field contribution to the free energy shown in Eq. (18) is given by:
\be
F_{mft} = NT((1-\tilde h)\ln(1-\tilde n_f) + \tilde h\ln(\tilde n_f) - (1 + \tilde w) \ln(1 + \tilde n_b) + \tilde w \ln(\tilde n_b)) + T \ln(\frac{\tilde n_b + \tilde n_f}{ \tilde n_b \tilde n_f}) 
\ee
yields the saddle-point equations:
\bea
\tilde n_f + \frac{1}{N}(1 - n_{\alpha})= \tilde h  \\
\tilde n_b + \frac{1}{N} n_{\alpha} = \tilde w 
\eea
where $n_{\alpha} = \frac{1}{e^{2\beta \zeta} + 1} = \frac{1}{\frac{\frac{1}{\tilde n_f} - 1}{\frac{1}{\tilde n_b} + 1} + 1}$ is the gaussian contribution to the $-\frac{1}{Q}<[\theta,\theta^{\dagger}]>$ term of the constraints which we need to include in the saddle-point (ie. at the mean field level).  With this assignment, we see that at the mean field level we recover the results given by the earlier Abrikosov pseudofermion and Schwinger boson approaches.  Analytic solution of these equations shows a recovery of the constraints for these two cases when the Young tableau is a column or a row respectively, while for Y = 0 we have in general a mixed case (see Fig. 3b).

\vskip1pc

(b) {\it{Abrikosov fermions.}}
The exact entropy (normalized) of a column of height $\frac{N}{2}$ and width 1, is given by:
\be
\frac{2S}{N} = \frac{2}{N}\ln\left(\begin{array}{c}N \\ \frac{N}{2} \end{array}\right) 
\ee
Using Stirling's approximation to expand the factorials, we obtain
\be
\frac{2S_{Stirling's}}{N}= 2\ln2 - \frac{1}{N}\ln(\frac{\pi N}{2})
\ee
which should be the best that a $\frac{1}{N}$ theory can do.  With self-consistent corrections, one might hope to approach the exact solution.  
  
For Abrikosov pseudofermions the mean field + gaussian corrections free energy is given by: 
\bea
F &= -NT \ln(1 + e^{-\beta \lambda}) - \lambda Q + \frac{T}{2}\ln(2 \pi N \tilde n_f (1- \tilde n_f)) \nonumber \\ &= NT\ln(1 - \tilde n_f) - NqT \ln(\frac{1 - \tilde n_f}{\tilde n_f}) + \frac{T}{2}\ln(2\pi N \tilde n_f (1 - \tilde n_f))
\eea
where the last term is the gaussian corrections to the mean field result.  
Taking the usual mean field saddle-point $\frac{\partial F_{mft}}{\partial n_f} = 0$ (mfc) we obtain $\tilde n_f = q$, such that the entropy (at $q = \frac{1}{2}$), 
\be
\frac{2S_{mfc}}{N} = 2\ln2 - \frac{1}{N}\ln(\frac{\pi N}{2})
\ee
reproduces the Stirling's approximation to the exact result. An attempt at a fully 'self-consistent' (fsc) description might set $\tilde n_f = \frac{q + \frac{1}{2N}}{1 + \frac{1}{N}}$ but such attempts do not seem to improve the accuracy.
\vskip1pc

{(c) {\it{Schwinger Bosons.}}}
For the simple case of a purely bosonic description of spins, the exact entropy (for a row of width $\frac{N}{2}$ and height 1) is given by:
\be
\frac{2S_{exact}}{N} = \frac{2}{N}\ln\left(\begin{array}{c}N + \frac{N}{2} - 1 \\ \frac{N}{2} \end{array}\right) 
\ee
which using Stirling's approximation simplifies to:
\bea
&\frac{2S_{Stirling}}{N} = (3 - \frac{1}{N})\ln(\frac{3}{2} - \frac{1}{N}) + \ln 2 - \frac{1}{N}\ln(\pi N) + (- 2+ \frac{1}{N})\ln(1 - \frac{1}{N})
\eea
which to leading order in N (ie dropping $\frac{1}{N}$ inside log's) yields
\be
\frac{2S_{Stirling,lead}}{N} = 3\ln(\frac{3}{2}) + \ln(2) - \frac{1}{N} \ln(\frac{3\pi N}{2})
\ee
The mean field + gaussian corrections free energy is given by
\bea
F &=& NT\ln(1 - e^{-\beta \lambda}) -\lambda Q + \frac{T}{2}\ln(2\pi N \tilde n_b(1 + \tilde n_b)) \nonumber \\ &=& -NT\ln(1 + \tilde n_b) - NqT\ln(\frac{1 + \tilde n_b}{\tilde n_b}) + \frac{T}{2}\ln(2\pi N \tilde n_b (1 + \tilde n_b))
\eea
For the usual mean field saddle-point, we obtain $\tilde n_b = q$ so the entropy is found to be:
\be
S_{mfc} = N (1+q)\ln(1+q) - Nq\ln(q) - \frac{1}{2}\ln(2\pi Nq(1+q))
\ee
which reduces to
\be
\frac{2S_{mfc}}{N} = 3\ln(\frac{3}{2}) + \ln2 - \frac{1}{N}\ln(\frac{3\pi N}{2}) 
\ee
for the special case $q = \frac{1}{2}$, agreeing with the leading approximation to Stirling's approximation to the exact entropy.  For a fully 'self-consistent' saddle-point the consistency conditions would yield $\tilde n = \frac{q + \frac{1}{2N}}{1 - \frac{1}{N}}$ which is less close to the known exact result.

\vskip1pc

{\centering{B. Comparison of Accuracy}\\}

\vskip1pc

{\centering{\it 1. Susy}\\}

\vskip1pc
\begin{figure}
\includegraphics[scale=0.5]{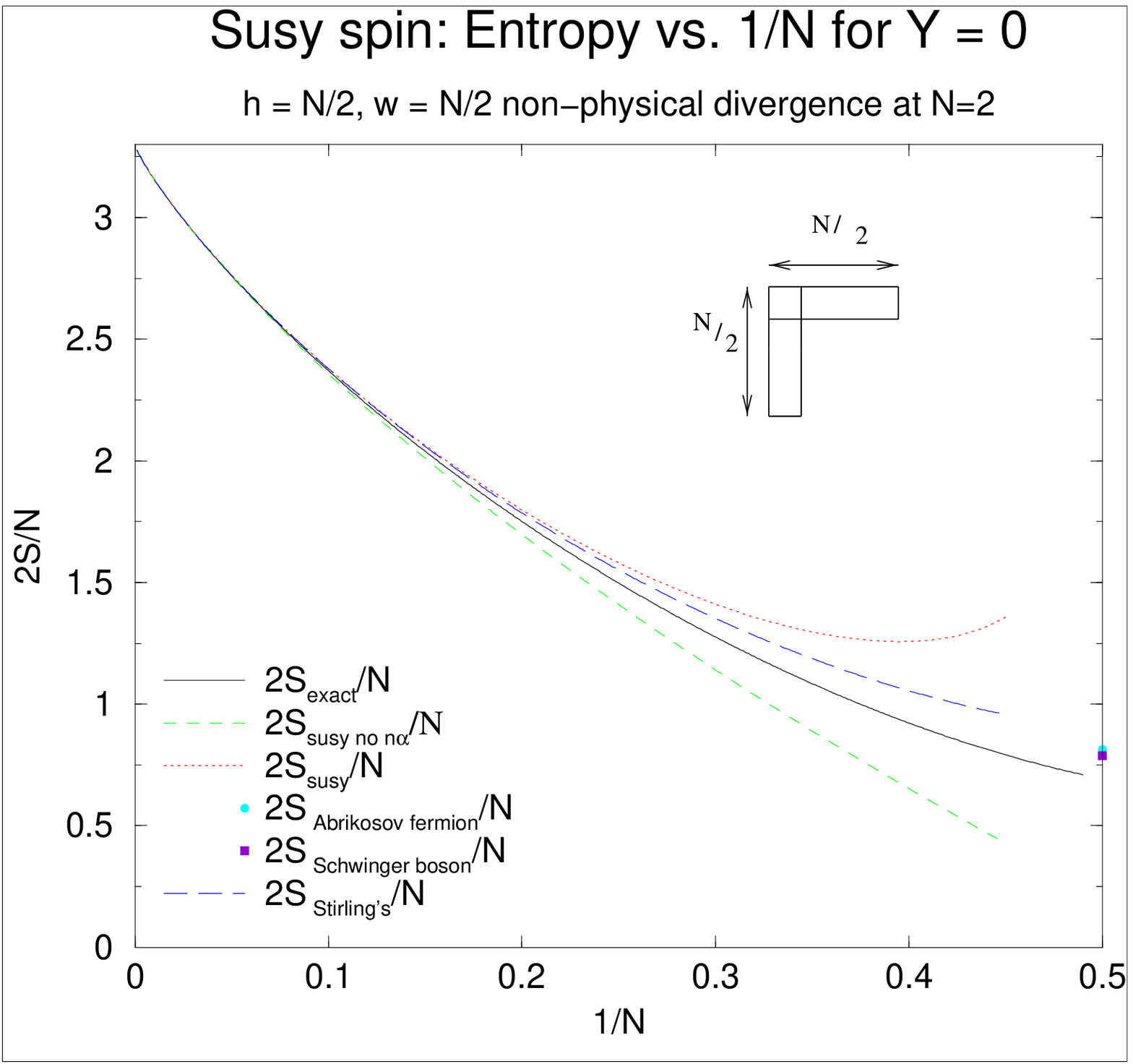}
\includegraphics[scale=0.5]{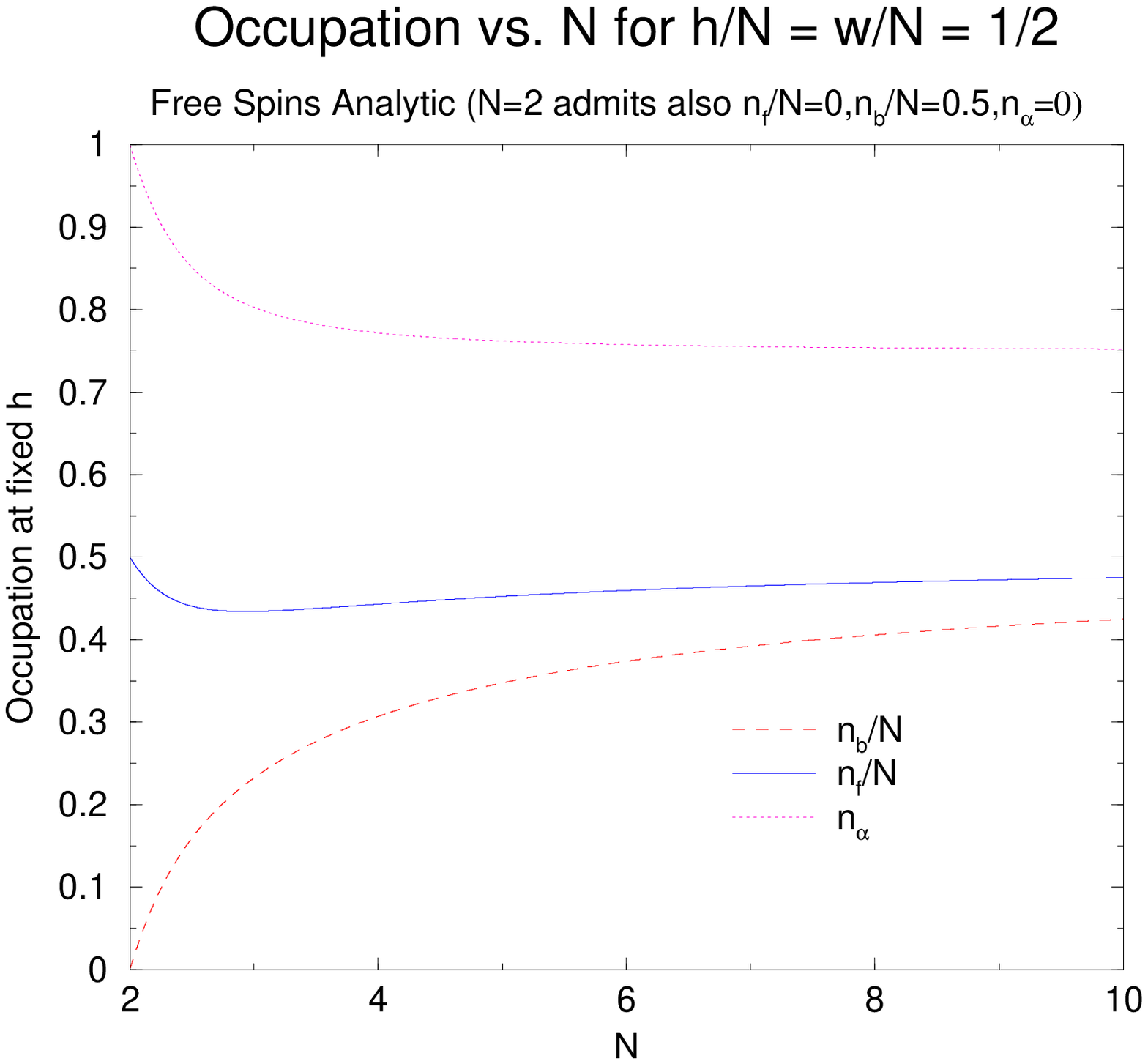}
\caption{\label{figure 3}a) As N increases the supersymmetric spin mean field + gaussian corrections at Y = 0 approaches the exact result, where the saddle-point has been chosen to maximize the number of terms kept at the mean field level. Choosing the incorrect saddle-point (ignoring the 1/N term in the constraint for Y) underestimates the entropy.  About the correct saddle-point the divergence as one approaches N=2 is a non-physical term arising from the gaussian fluctuations in $\zeta$ and $\lambda$ and might be expected to be cancelled by fluctuations in the corner box. Note that the correct saddle-point reproduces the Stirling's approximation when sufficiently far from the non-physical divergence at N=2.  At N=2, Schwinger bosons and Abrikosov fermions solutions are found, but the entropy has an additional non-physical $-\frac{1}{2}\ln(4\pi 0)$ term.  b) In the limit N$\rightarrow$2, either Schwinger bosons or Abrikosov fermions are produced by the mean field conditions.  For N$>$2 there is a unique analytic solution at Y=0 for $\tilde h=1/2$}
\end{figure}
At N=2, the susy approximation reverts to either Abrikosov fermions or Schwinger bosons (see Fig 3(b)) recovering the entropy of these approximations if one is willing to remove a divergent term by hand: $-\frac{1}{N} ln(2\pi N (0))$.  It is interesting to speculate that a correction of the form $-\frac{2}{N^2} log(\frac{\tilde n_b (1 + \tilde n_b) + \tilde n_f (1 - \tilde n_f)}{2 \pi \tilde n_b (1 + \tilde n_b) \tilde n_f (1 - \tilde n_f)})$ may be necessary to remove this divergence in these fluctuations and might arise in higher order diagrams, but such a correction is beyond the scope of this paper.  Nonetheless, the agreement for N$>$2 is quite reasonable (see Fig. 3(a)).

\vskip1pc

{\centering{\it{2. Abrikosov fermions}}\\}

\vskip1pc

To compare these approximate solutions with the exact solution, by $N = 10$ a graph is indistinguishable by the eye (see Fig. 4(a)).  At $N = 2$, $\frac{2S_{exact}}{2} = \ln 2$, so the percentage error is given by 
\be
\frac{\frac{2S_{Stirling's,mfc}}{2} - \frac{2S_{exact}}{2}}{\frac{2S_{exact}}{2}} * 100 = 17 \% 
\ee
which agrees with the fully self-consistent version for $q =\frac{1}{2}$. 
So there seems to be no advantage here to fixing $\frac{\partial F_{fsc}}{\partial \tilde n_f} = 0$ over $\frac{\partial F_{mft}}{\partial \tilde n_f} = 0$.

\begin{figure}
\includegraphics[scale=0.5]{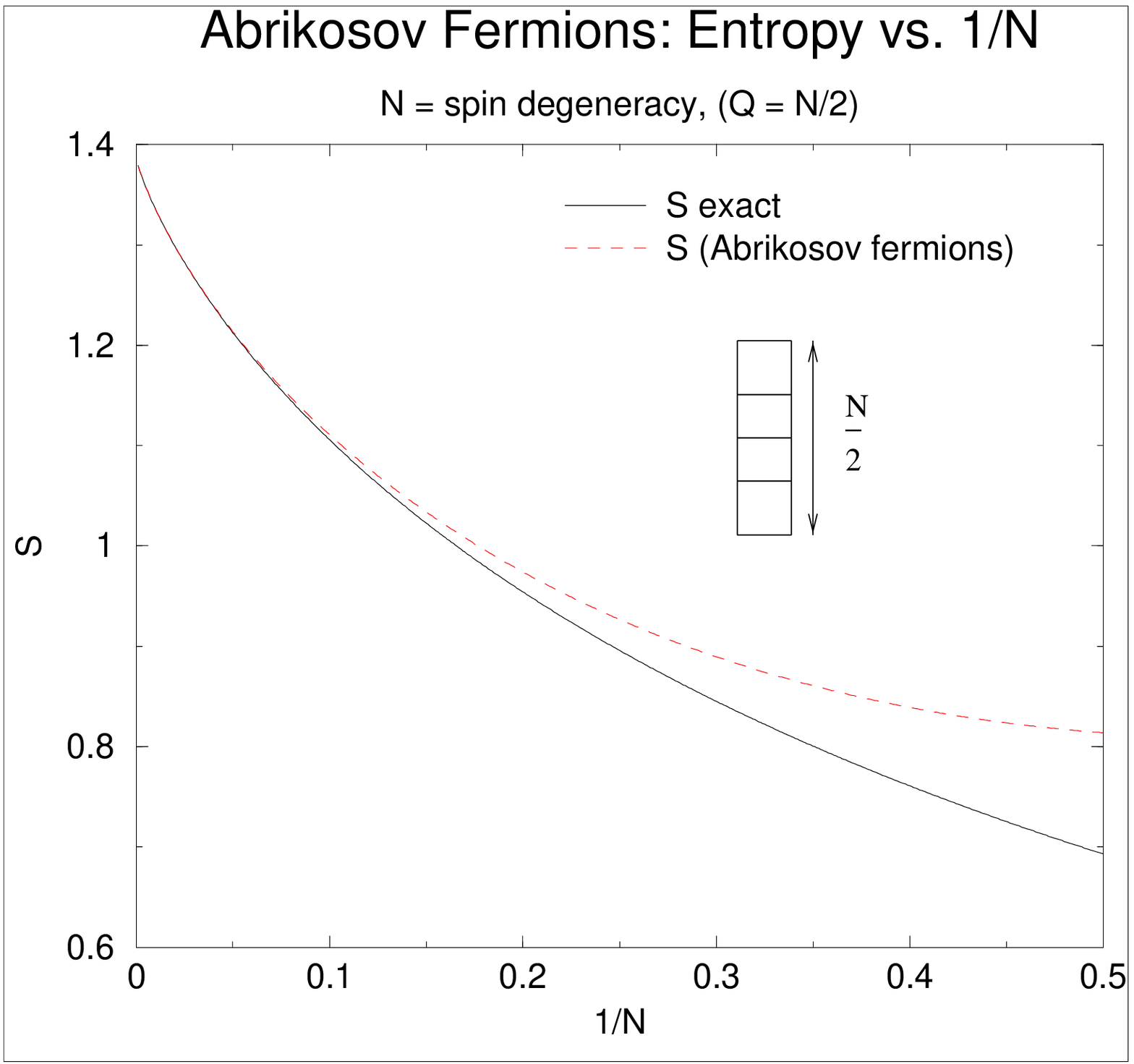}
\includegraphics[scale=0.5]{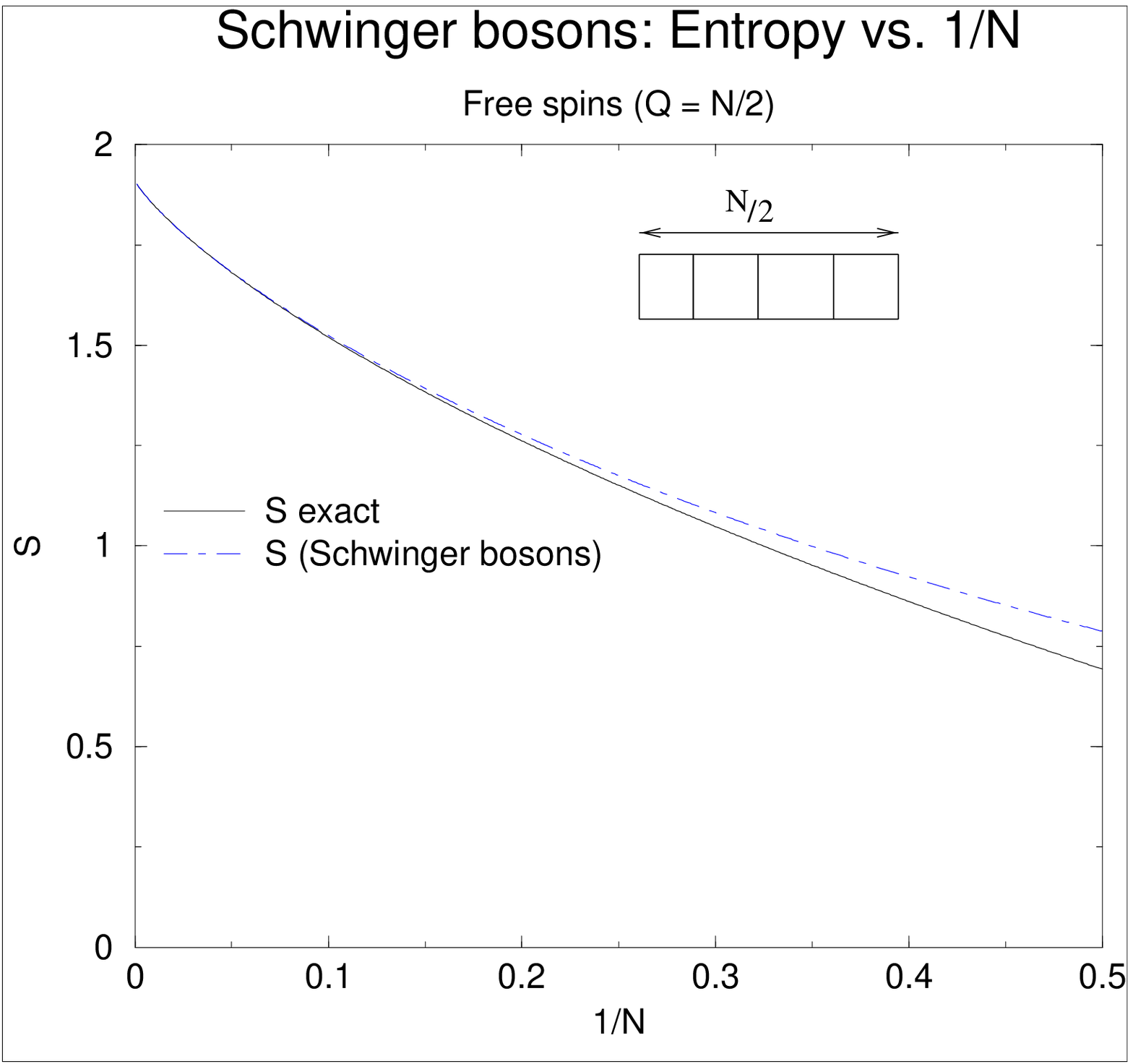}
\caption{\label{figure 4}Beyond N$\approx$ 12, Abrikosov fermion and Schwinger boson solutions for the entropy of a free spin are indistinguishable to the eye from the exact for a free spin.  Here we compare the exact entropy as a function of spin degeneracy N two large N approximations. For Abrikosov fermions, at the mean field saddle-point, mean field + gaussian corrections results agree with Stirling's approximation to the exact result and an attempt to introduce self-consistency. The slave boson mean field saddle-point out-performs both the full Stirling's approximation and an attempt at full self-consistency.}
\end{figure}

\vskip1pc

{\centering{\it{3. Schwinger bosons}}\\}

\vskip1pc

The self-consistent approximation ($\frac{\partial F_{fsc}}{\partial n_b} = 0$) does dramatically worse ($-40\%$ error at N =2) than mean field solution ($\frac{\partial F_{mft}}{\partial n_b} = 0$).  At $N=2$, 
\be
\frac{\frac{2S_{mfc}}{2} - \frac{2S_{exact}}{2}}{\frac{2S_{exact}}{2}} * 100  = 14 \% 
\ee
whereas

\be
\frac{\frac{2S_{Stirling's}}{2} - \frac{2S_{exact}}{2}}{\frac{2S_{exact}}{2}} * 100  = 17 \% 
\ee
while for large N, both approximations approach the correct result, above $N=12$ being indistinguishable (see Fig.4(b)).  It is interesting to note that the leading $\frac{1}{N}$ approximation to Stirling's formula does a better job than the formula itself, and the mfc approach reproduces this nice feature at N=2.

\vskip1pc

{\centering{{3. ATOMIC MODEL}\\}}
  
\vskip1pc
Next, we consider a simple atomic model.  In such a model there is an excitation energy for a spin to leave its localized site, creating an empty or vacuum site.  The Hamiltonian for such a system is particularly simple, being given by simply the energy difference between the unoccupied site and the occupied spin-ful site.  In terms of Hubbard operators, this can be written:
\be
H = -E_d X_{0 0}.
\ee

\vskip1pc

{\centering{A. Entropies, Heat Capacities and Wavefunctions}\\}

\vskip1pc

{\centering{\it{1. Exact Results}}\\} 

\vskip1pc

(a) {\it{N=2}}.
For a single SU(2) spin with no double occupancy, we know that the exact partition function is given by
\be
Z = 2 + e^{\beta E_d}
\ee
which means that the free energy is given by
\be
F = -T \ln(2 +  e^{\beta E_d})
\ee
the entropy is 
\be
S = \ln(2 +  e^{\beta E_d}) + \frac{ -E_d e^{\beta E_d}}{T(2 +  e^{\beta E_d})}
\ee
and the heat capacity is given by 
\be
C_v = \frac{2 (\beta E_d)^2 e^{-\beta E_d}}{(2 e^{-\beta E} + 1)^2}
\ee
By writing out the possible states of the system and their energies, we can count the total number of states available per site, as done for the Q = 1, Y = 0, N = 2 case above.  This procedure allows us to compute exact results for thermodynamic quantities for larger representations.  In Table I, we list the results for Q=2,3 calculated in Appendices B, C. 
\begin{table}[hbtp]
\begin{tabular}{llll}
$Q,Y $&$ F $&$ S = - \frac{dF}{dT} $& $C_v = -T\frac{d^2F}{dT^2}$ \\
\hline
$\{2,1\} $&$-T\ln(1 + 2e^{\beta E_d} + e^{2\beta E_d})$ &$ \ln(1
+ 2e^{\beta E_d} + e^{2\beta E_d}) + \frac{-2 \beta E_d}{e^{-\beta E_d} + 1}
$&$ \frac{2(\beta E_d)^2 e^{-\beta E_d}}{(1 + e^{-\beta E_d})^2}   $ \\
$\{2,-1\} $&$ -T\ln(3 + 2e^{\beta E_d}) $&$ \ln(3 + 2e^{\beta E_d}) +
\frac{-2\beta E_d}{3 e^{-\beta E_d} + 2} $&$ \frac{6(\beta E_d)^2e^{-\beta
E_d}}{(3 e^{-\beta E_d} + 2)^2}   $ \\
$\{3,0\}$ & $-T\ln2 -T\ln(1 + 2e^{\beta E_d} + e^{2\beta E_d}) $&$ \ln2 +
\ln(1 + 2e^{\beta E_d} + e^{2\beta E_d}) + \frac{-2 \beta E_d}{e^{-\beta E_d} +
1} $&$ \frac{2(\beta E_d)^2 e^{-\beta E_d}}{(1 + e^{-\beta E_d})^2}   $ \\
$\{3,-2\} $&$ -T\ln(4 + 3e^{\beta E_d}) $&$ \ln(4 + 3e^{\beta E_d}) +
\frac{-3 \beta E_d}{4 e^{-\beta E_d} + 3} $&$ \frac{12(\beta E_d)^2 e^{-\beta
E_d}}{(4e^{-\beta E_d} + 3)^2}   $ \\ $\{3,2\} $&$  -T\ln(1 + 2e^{\beta
E_d} + e^{2\beta E_d}) $&$ \ln(1 + 2e^{\beta E_d} + e^{2\beta E_d}) +
\frac{-2 \beta E_d}{e^{-\beta E_d} + 1} $&$ \frac{2(\beta E_d)^2 e^{-\beta
E_d}}{(1 + e^{-\beta E_d})^2} $
\end{tabular}
\caption{\label{Table I} Free energy, entropy and heat capacity values of {Q, Y} restricted to the case when N = 2.  The first non-trivial case with Y=0 has Q = 3. }  \end{table}

\vskip1pc

{(b) {\it{Exact Wavefunctions.}}
For the simple case (Q=1, N=2) we can choose to represent the spin as either a fermion or a boson, as in reality a single spin is neither.  This also applies to the empty state.  It can be described by a grassman variable or a spinless boson.  This dictates the possible wavefunctions for Q = 1, Y = 0, N = 2 to be:
\be
f^{\dagger}_{\uparrow}|0>, \hspace{2mm} f^{\dagger}_{\downarrow}|0>,\hspace{2mm}b^{\dagger}_{\uparrow}|0>, \hspace{2mm} b^{\dagger}_{\downarrow}|0>,\hspace{2mm}\phi^{\dagger}|0>, \hspace{2mm} \chi^{\dagger}|0>
\ee
so we have a double-counting of the available states of the system.  We have four states of energy 0, and two of energy -E$_d$.  To agree with physical results, we remove this double counting, so for each partition function written we divide the number of available states to the system by two.  That is, for the single box case rather than $Z = 4 + 2e^{\beta E_d}$ as one might expect from the degeneracies of the wavefunctions written, we will have $Z = 2 + e^{\beta E_d}$ which agrees with our prior knowledge of this system.

The rules for writing wavefunctions are fairly simple.  Since 
\be
[\theta^{(\dagger)}, Q] = [\theta^{(\dagger)}, Y] = 0 \Rightarrow \forall |\nu> \hspace{2mm} \in \{Q = Q_0, Y = Y_0\} \exists \theta^{(\dagger)}|\nu> \in \{Q = Q_0, Y = Y_0\}
\ee
\bea
[\theta^{(\dagger)}, X_{\sigma \sigma'}] = 0 &\Rightarrow& [\frac{1}{Q}\{\theta \theta^{\dagger} - \theta^{\dagger} \theta\}, X_{\sigma \sigma'}] = 0 \cr &\Rightarrow& \forall |\nu> \hspace{2mm} \in \{Q = Q_0, Y = Y_0\} \exists X_{\sigma \sigma'}|\nu> \in \{Q = Q_0, Y = Y_0\}  
\eea
\bea
\{\theta^{(\dagger)}, X_{0 \sigma}\} = 0 &\Rightarrow& [\frac{1}{Q}\{\theta \theta^{\dagger} - \theta^{\dagger} \theta\}, X_{0 \sigma}] = 0 \cr &\Rightarrow&\forall |\nu> \hspace{2mm} \in \{Q = Q_0, Y = Y_0\} \exists X_{0 \sigma}|\nu> \in \{Q = Q_0, Y = Y_0\} 
\eea
Where we have used the fact that the Hubbard operators conserve particle number and slave representation in the later two statements.  This means that by finding a `highest weight' state and applying these operators we can generate a basis of states consistent with $Q = Q_0, Y = Y_0, N = N_0$.  An additional subtlety to be aware of is that the Hubbard commutation relations mean that for blind application of this method we will end up with a non-orthogonal basis since:
\be
[X_{\uparrow \downarrow}, X_{0 \uparrow}] = - X_{0 \downarrow}
\ee
so we will intentionally discard one of the three to form a basis.

\vskip1pc

{(c) {\it{Exact results at general N.}}}
To calculate the general partition function for the atomic model, we need to simply count the number of possible states of the system at each energy.  This can be simply performed by writing the corresponding Young Tableau for each energy level and using the counting shown above in Eq. (8) (see Fig. 5).  As one can only have one slave fermion (as it is a grassman variable), but up to h slave bosons, an L-shaped tableau of energy 0 and dimensions (h,w) allows two L-shaped energy $-E_d$ states to arise after application of the Hubbard operators with corresponding dimensions (h, w-1) and (h-1, w).  Application of Eq. (8) yields the coefficient of the $e^{\beta E_d}$ term.  At energy $-E_d$ we have Young Tableaux with dimension (h-1,w-1) and (h-2,w), and one can continue this procedure until energy $- h E_d$ when only one diagram (1,w-1) remains.  This procedure allows us to generate exact results for the free energy in cases that would otherwise prove unaccessible to us. 
 
\begin{figure}
\includegraphics[scale=0.8]{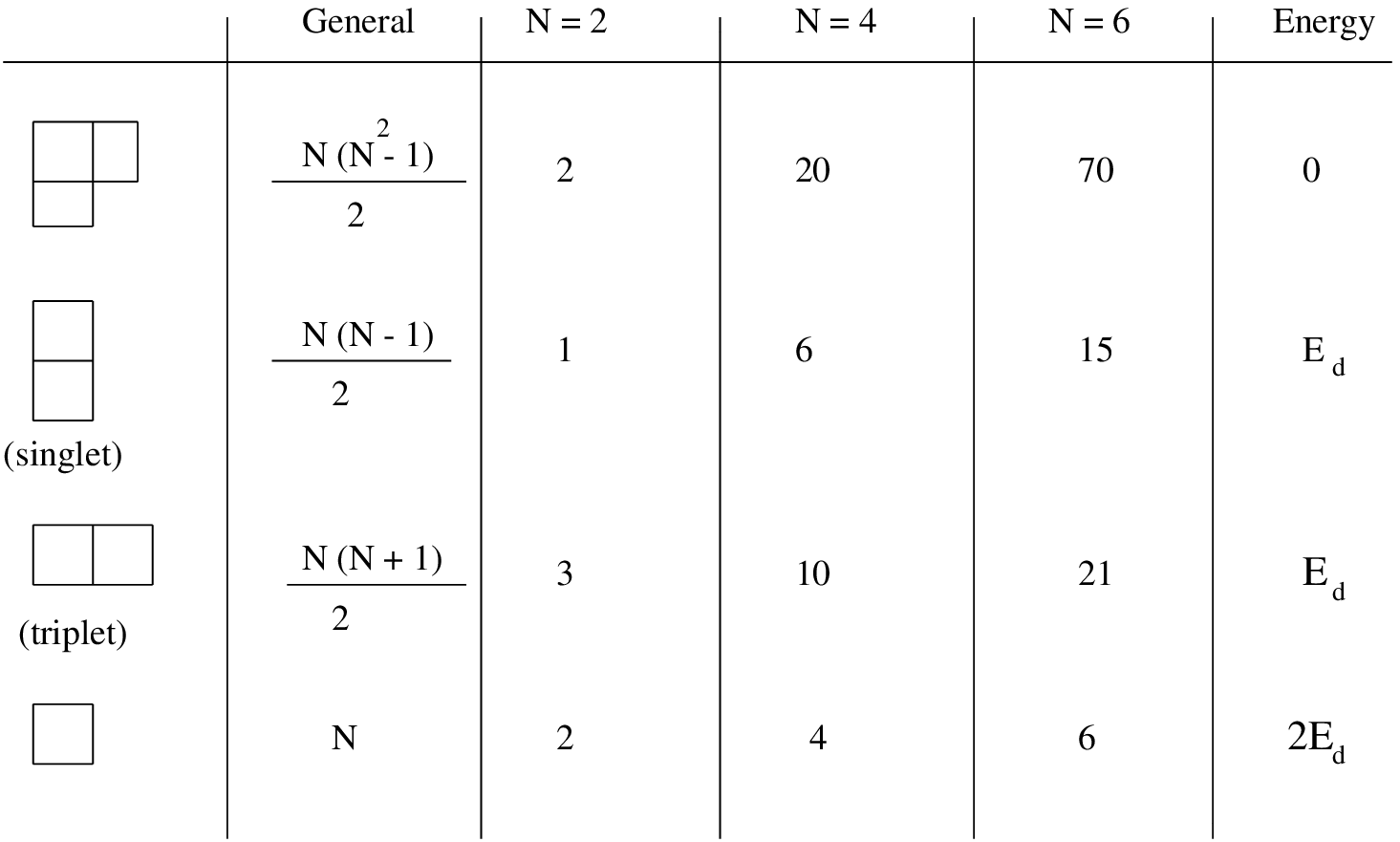}
\caption{\label{figure 5}Illustration of counting procedure to generate the partition function of Q=3,Y=0. }
\end{figure}
Using the above algorithm, it is simple to generate exact results for the free energy of an arbitrarily shaped Young Tableau (see Table II).

\begin{table}[hbtp]
\begin{tabular}{ll}
$Q,N $&$ F $ \\
\hline
$\{3,4\}$ & $-T\ln(4) - T\ln(5+4e^{\beta E_d} + e^{2\beta E_d})$\\$\{5,6\} $&$-T\ln(21) -T\ln(16 + 15e^{\beta E_d}+6e^{2\beta E_d} + e^{3\beta E_d})$ \\ $\{7,8\}$ & $-T\ln(120) - T\ln(55+56e^{\beta E_d}+28e^{2\beta E_d} + 8e^{3\beta E_d} + e^{4\beta E_d})$\\$\{9,10\}$ & $-T\ln(715) - T\ln(196 + 210e^{\beta E_d} + 120 e^{2\beta E_d} + 45 e^{3\beta E_d} + 10 e^{4\beta E_d} + e^{5\beta E_d})$ 
\end{tabular}
\caption{\label{Table II}The Y=0 series with $\tilde h = \frac{1}{2}$. Using the above algorithm, it is simple to generate exact results for the free energy of an arbitrarily shaped Young Tableau.}  \end{table}

{\centering{\it{2. Mean Field + Gaussian corrections}}\\}

\vskip1pc
{(a) {\it{Susy.}} }
For the full atomic model, the (mean field) Hamiltonian reads
\bea
H = -E_d (\hat n_{\phi} + \hat n_{\chi}) + \lambda(\hat n_b + \hat n_f + \hat n_{\chi} + \hat n_{\phi} - Q_0) + \zeta (\hat n_f + \hat n_{\phi} - (\hat n_b + \hat n_{\chi}) + \frac{1}{Q_0}<[\theta,\theta^{\dagger}]> - Y_0)
\eea
where $n_{\chi}$ and $n_{\phi}$ are slave fermion and slave boson fields respectively.  As we saw in the free spin limit, it is useful to express Q$_0$ and Y$_0$ in terms of the height and width of the Young tableau (h,w), which slightly modifies the constraint on Q$_0$, as we absorb a term $\frac{\lambda}{N}$ into the gaussian fluctuations.  Then the free energy is

\be
F = F_{mft} + \frac{1}{N}F_{g.c}
\ee
where 
\bea
F_{mft} &=& -T \ln(1 + e^{-\beta (\lambda_b - E_d)}) - NT \ln(1 + e^{-\beta \lambda_f}) + T\ln(1 - e^{-\beta(\lambda_f - E_d)}) + NT\ln(1-e^{-\beta \lambda_b}) + T\ln(e^{\beta \zeta} + e^{-\beta \zeta}) \nonumber \\& & - \lambda (h + w - 1) - \zeta (h - w) \nonumber \\ &=& NT(\tilde h \ln(\tilde n_f) + (1 - \tilde h)\ln(1 - \tilde n_f) + \tilde w \ln(\tilde n_b) - (1 + \tilde w) \ln(1 + \tilde n_b)) \nonumber \\ & & + T\ln(1 - n_{\chi}) - T\ln(1 + n_{\phi}) - T\ln(\frac{\tilde n_b \tilde n_f}{\tilde n_b + \tilde n_f})
\eea
and
\bea
\frac{1}{N}F_{g.c} &=& F_{\delta \lambda_b} + F_{\delta \lambda_f}  \nonumber \\  &=& + \frac{T}{2}\ln((2 \pi)^2 ((N\tilde n_b(1 + \tilde n_b) + n_{\chi}(1 - n_{\chi})-n_{\alpha}(1-n_{\alpha}))(N\tilde n_f (1 - \tilde n_f) + n_{\phi}(1 + n_{\phi})-n_{\alpha}(1-n_{\alpha})) \nonumber \\ & &- n_{\alpha}^2 (1-n_{\alpha})^2))
\eea
where $\tilde n_b = \frac{1}{e^{\beta \lambda_b} - 1}$ and $\tilde n_f = \frac{1}{e^{\beta \lambda_f} + 1}$.  We have again dropped the $T\ln2$ from the double counting of the corner box and the familiar $- T\ln(\frac{\tilde n_b \tilde n_f}{\tilde n_b + \tilde n_f})$ has again been included in the mean field to get us as close as possible to the correct saddle-point.  The saddle-point solution  $\frac{\partial F_{mft}}{\partial \lambda_f} = \frac{\partial F_{mft}}{\partial \lambda_b} = 0$ now yields: 
\bea
&\tilde n_f + \frac{1}{N} n_{\phi} + \frac{1}{N} (1 - n_{\alpha}) = \tilde h \\ &\tilde n_b + \frac{1}{N}n_{\chi} + \frac{1}{N} n_{\alpha} = \tilde w 
\eea

where $n_{\alpha} = \frac{1}{e^{2\beta \zeta} + 1} = \frac{1}{\frac{\frac{1}{\tilde n_f} - 1}{\frac{1}{\tilde n_b} + 1} + 1}$ gauges the degree by which a representation is fermionic or bosonic (refer to Appendix D for a discussion), $n_{\phi} = \frac{1}{e^{-\beta E_d} (\frac{1}{\tilde n_f} - 1) - 1}$ and $n_{\chi} = \frac{1}{e^{-\beta E_d} (\frac{1}{\tilde n_b} + 1) + 1}$.  Hence as in the free spin limit, we include as much as possible in the mean field saddle-point before treating the gaussian fluctuations.

As in the free spin case, the saddle-point has an analytic solution for columns or rows.  The special case Q =1, N=2 is of interest despite the non-physical divergence of the entropy noted in the free spin limit (Fig.3).  The heat capacity does not suffer from this malady, and is accessible to us as we can analytically (see Table III) remove the divergent terms in the entropy (which are independent of temperature--at least in the slave boson/slave fermion limits) by hand. Furthermore, the phase diagram generated by the saddle point solutions is interesting in its own right, with three possible solutions arising: slave bosons, slave fermions or mixed solutions.  In Fig.6 we sketch the simple picture arising.  By minimizing the free energy (see Fig. 6(a)), one is driven to the cross-over from a slave fermion description  to a slave boson description as one increases the value of $E_d$.  This appears to be the physically realized situation.  A more compelling phase diagram corresponds to maximizing the free energy of the saddle-point solutions, yielding a localized spin slave boson description, a small window of mixed solution followed by a slave fermion description at high temperatures which looks superficially similar to that seen close to quantum critical points (see Fig. 6(b)).  Unfortunately it is not as simple to remove the non-physical divergence from this term, and it appears that simple procedures would lead to a non-physical heat capacity, such that the mixed solution likely is not chosen in the atomic model.  It is interesting to note that at the mean field level the entropies match at these phase boundaries, and further that the difference between the entropy difference calculated from these approximations at the level of gaussian fluctuations would lead to a weakly positive heat capacity, leading one to believe that if the gaussian fluctuations were properly controlled in the supersymmetric approximation, one might in fact see a second order transition.  Unfortunately as yet we have not discovered a way to realize this goal.  It would be interesting however to imagine turning on a hybridization to a conduction sea and investigating if beyond some hybridization this state becomes realizable.  For general Young Tableau of the row type we see that an analytic slave fermion solution exists, the column type has an analytic slave boson solution arising from the mean field consistency conditions.  For more general L-shapes a numerical root-finding is necessary.

\begin{figure}
\includegraphics[scale=1.0]{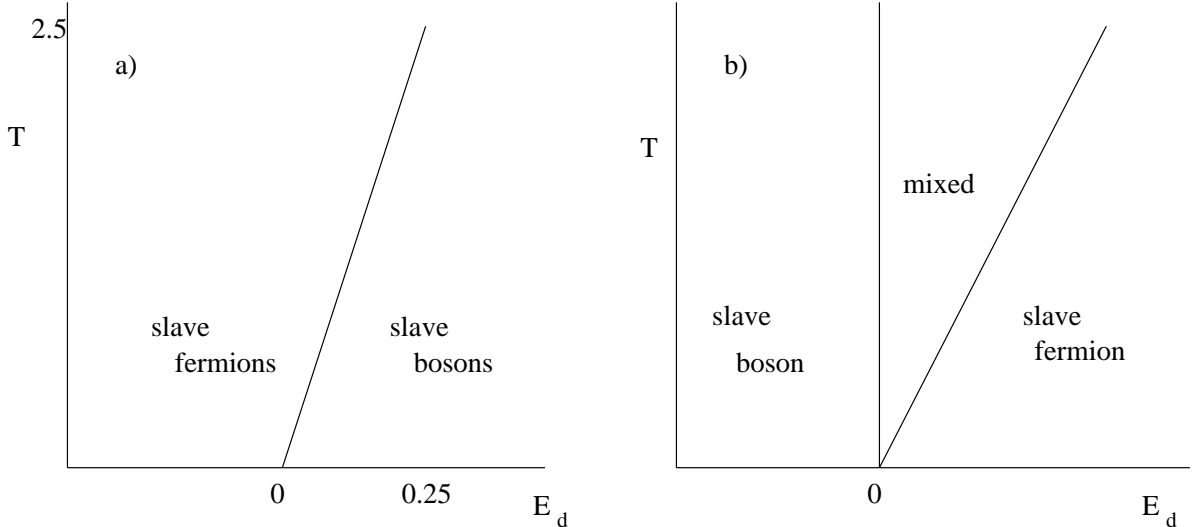}
\caption{\label{figure 6}The analytic saddle-point solutions at N = 2 allow two possible phase diagrams: a) The physical saddle-point which picks the smallest free energy of the saddle-point solutions. b) A non-physical saddle-point which picks the highest free energy.  While this looks qualitatively similar to that found close to a quantum critical point, we must remember that this model has no hybridization term.  Non-physical divergences in a) are simple to remove analytically (constant in S) and do not affect the heat capacity, but this is less clear in the mixed phase of b)--simple attempts to remove this divergence seem to lead to a non-physically negative heat capacity.  By turning on the hybridization, one might hope to garner a physical solution underlying this symmetry of the underlying formalism.}
\end{figure}
\begin{table}[hbtp]
\begin{tabular}{lllll}
$\tilde n_f $&$ \tilde n_b $&$ n_{\chi} $& $n_{\phi}$ &$n_{\alpha}$ \\
\hline
$0$ & $\frac{e^{-\beta \frac{E_d}{2}} (\sqrt{8 + 9 e^{-\beta E_d}} - e^{-\beta \frac{E_d}{2}})}{4 (1 + e^{-\beta E_d})}$&$\frac{2 + 3 e^{-\beta E_d}-e^{-\beta E_d/2}\sqrt{8 + 9 e^{-\beta E_d}}}{2 (1+ e^{-\beta E_d})} $&$0$&$0$\\$\frac{e^{-\beta E_d}}{2(1+e^{-\beta E_d})}$&$0$&$0$&$\frac{1}{1+e^{-\beta E_d}}$&$1$\\ $\frac{2 e^{-\beta E_d} - 1}{2(e^{-\beta E_d} + 1)}$& $\frac{1-e^{-\beta E_d}}{1 + e^{-\beta E_d}}$&$\frac{1-e^{-\beta E_d}}{1 + e^{-\beta E_d}}$&$\frac{2 e^{-\beta E_d} - 1}{(e^{-\beta E_d} + 1)}$&$\frac{2(2 e^{-\beta E_d} - 1)}{1 + e^{-\beta E_d}}$
\end{tabular}
\caption{\label{Table III}Three solutions at N =2: slave fermion, slave boson, mixed.  The latter only exists in the range $E_d = 0$ to $E_d = T\ln(2)$, has a constant ratio of slave partner to spin partner, and tunes continuously from slave fermion to slave boson. These analytic solutions allow us to obtain analytic solutions for the free energy, removing by hand non-physical divergences at N=2 for the slave boson and slave fermion limits. }  \end{table}

\vskip1pc
{(b) {\it{Slave fermions.}}}
For slave fermions, the Hamiltonian is
\be
H_{sf} = -E_d \hat n_{\chi} + \lambda((\hat n_{\chi} + \hat n_{b}) - Q_0)
\ee

such that the mean field free energy is:
\be
F_{mft} = -T\ln(1 + e^{-\beta(\lambda - E_d)}) + NTq\ln(\tilde n_b) - NT(1 + q)\ln(1 + \tilde n_b)
\ee
and the gaussian corrections to this are
\be
\frac{1}{N}F_{g.c} =\frac{T}{2}\ln(2\pi (N\tilde n_b(1 + \tilde n_b) + n_{\chi} (1 - n_{\chi})))
\ee

where $n_{\chi} = \frac{1}{e^{-\beta E_d}(1 + \frac{1}{\tilde n_b}) + 1}$ is the number of slave fermions, and the particles have been constrained such that 

\be
\tilde n_b + \frac{1}{N} n_{\chi} = q
\ee

For this case, we see that as $T \rightarrow 0$ for $E_d < 0$, $n_{\chi} \rightarrow 0$ and $\tilde n_b \rightarrow q$, while for $E_d > 0$, we have $n_{\chi} \rightarrow 1$ and $\tilde n_b \rightarrow q - \frac{1}{N} n_{\chi}$ (see Appendix D).

\vskip1pc

{(c) {\it{Slave bosons.}}}
For slave bosons, the Hamiltonian is given by

\be
H_{sb} = -E_d \hat n_{\phi} + \lambda(\hat n_f + \hat n_{\phi} - Q_0)
\ee
which means that the mean field free energy is  
\be
F_{mft} = T\ln(1 - e^{-\beta(\lambda - E_d)}) + NT(1 - q)\ln(1 - \tilde  n_f) + NT q \ln(\tilde n_f)
\ee
and the gaussian corrections are
\be
\frac{1}{N}F_{g.c} = \frac{T}{2}\ln(2\pi(N \tilde n_f(1 - \tilde n_f) + n_{\phi}(1+ n_{\phi})))
\ee
where $n_{\phi} = \frac{1}{e^{-\beta E_d}(\frac{1}{\tilde n_f} - 1) - 1}$ and the constraint maintained is:
\be
\tilde n_f + \frac{1}{N} n_{\phi} = q
\ee

For $E_d>0$, as $T\rightarrow 0$, $\tilde n_f \rightarrow 0$ and $\frac{1}{N} n_{\phi} \rightarrow q$, while for $E_d<0$, $n_{\phi} \rightarrow 0$ and $\tilde n_f \rightarrow q$.    Despite the complete transition to the bosonic vacuum state for $E<0$, at low T some entropy does remain in the system.  Both entropies remain positive in this approximation, while the entropy of the system with $E>0$ is much larger.

\vskip1pc

{\centering{B. Comparison of accuracy}\\}

\vskip1pc

{\centering{\it{1. N=2, q = $\frac{1}{2}$}}\\}

\vskip1pc

As these are all large N techniques, we note that this is an especially cruel limit.  Nonetheless, a comparison shows that the heat capacity of slave bosons does the best job for $E_d>0$ (empty state favored) while the heat capacity for slave fermions does the best job for $E_d<0$ (spin state favored).  Choosing the lowest free energy saddlepoint corresponding to Fig 6 (a), the susy approximation agrees with the slave boson result for $E_d>0$ and slave fermion for $E_d<0$ at low temperatures (see Fig. 7). 

\begin{figure}
\includegraphics[scale=0.5]{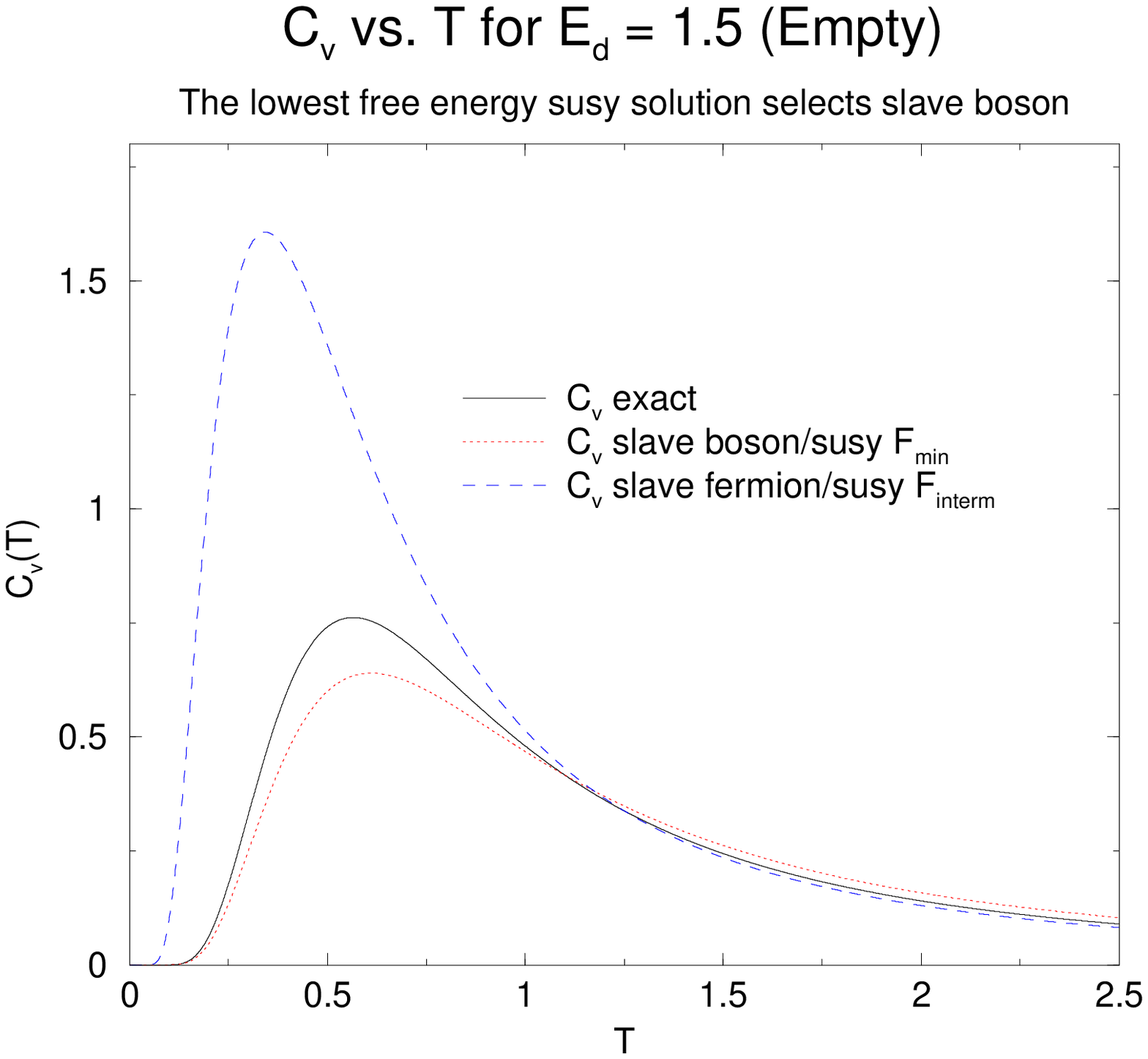}
\includegraphics[scale=0.5]{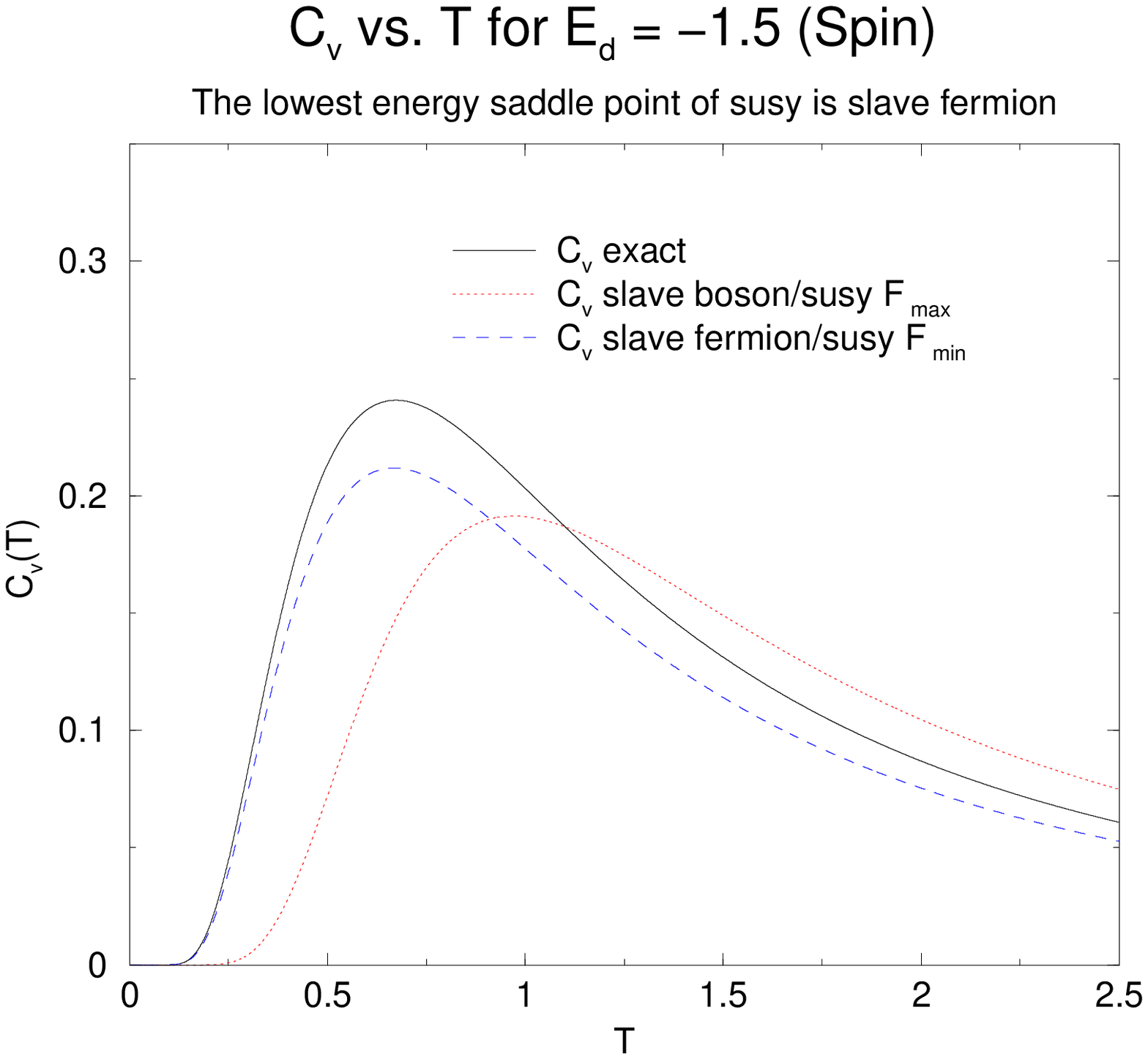}
\caption{\label{figure 7}(a) C$_v$(T) for $E_d=1.5$, Q = 1, N=2;(b) C$_v$(T) for $E_d=-1.5$, Q = 1, N=2. When evaluated at the mean field saddle-point, susy is able to describe both the spin and charge dominated regions of the atomic model.  A non-physical divergence can be removed by hand for the lowest energy saddle point solutions and does not affect the heat capacity.  The potentially interesting mixed saddle-point solution does not have this nice feature and has been omitted as uncontrolled.}
\end{figure}

{\centering{\it{2. The Y=0 series -- results and errors}}\\} 

\vskip1pc
In Fig. 8 we compare the heat capacity of our approximation with the exact result.  We see that as we increase N while holding $\tilde h = \frac{1}{2}$ and Y=0 constant the fit improves, although it is surprisingly good already at N=4.  This is partly because the non-physical divergence of the susy approximation as $N \rightarrow 2$ does not affect the heat capacity as noted in section A2 (a). 
\begin{figure}
\includegraphics[scale=1.0]{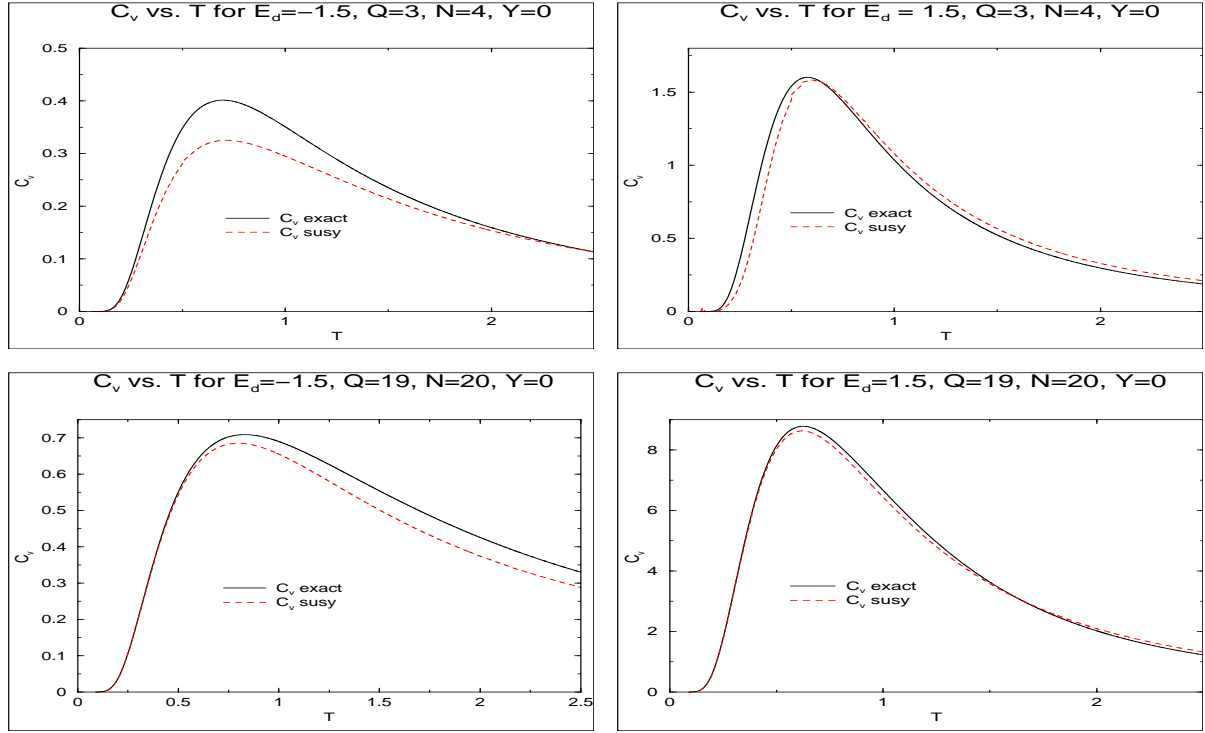}
\caption{\label{figure 8}Heat capacity as N increases.}
\end{figure}
In Fig. 9 we see that the entropy being an extrinsic quantity improves dramatically in its description from N=3 to N=20 for the same Y=0 series.
\begin{figure}
\includegraphics[scale=1.0]{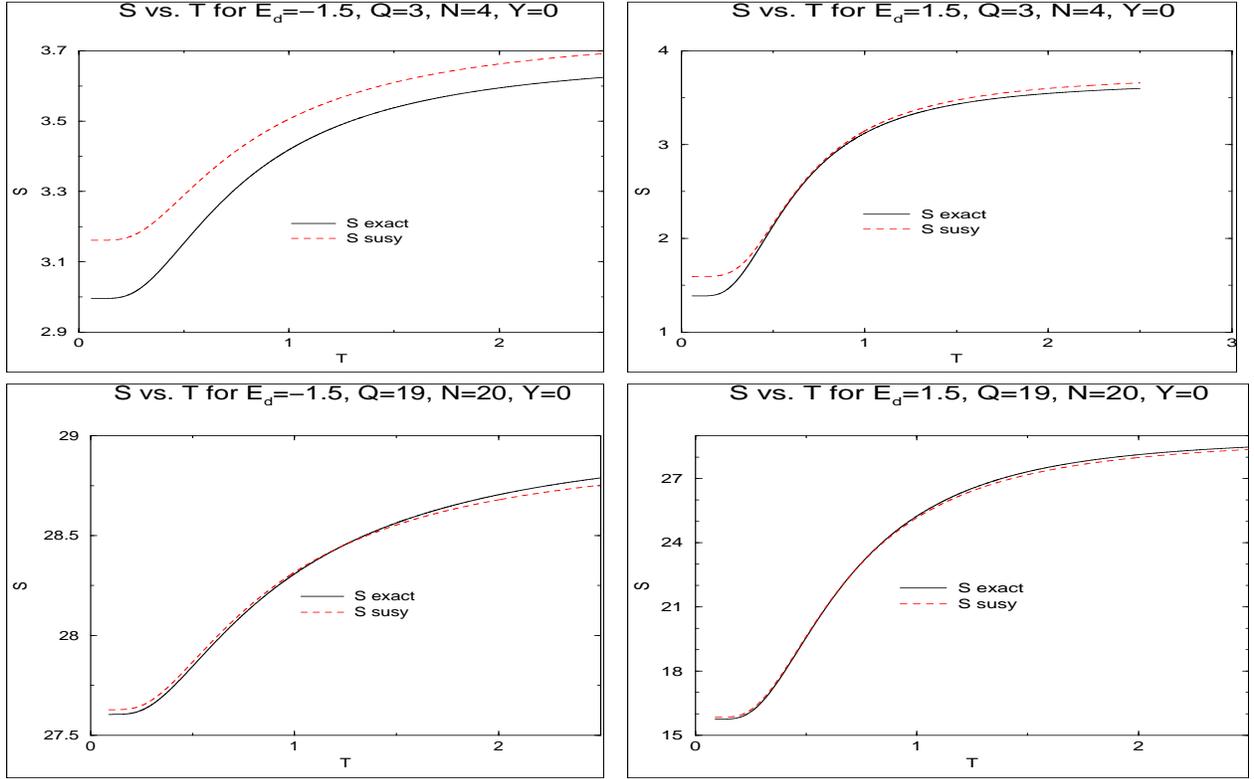}
\caption{\label{figure 9}Entropy as a function of N. Since entropy scales with N, at large values it is approximately exact. }
\end{figure}

But the eye is somewhat deceiving.  In Fig. 10 and 11 we show the absolute errors the entropy and heat capacity respectively as a function of temperature.  Both results seem to indicate that there is an error inherent to the approximation of the order of $\frac{1}{N}$, which may also be present for the slave boson and slave fermion approximations and should be investigated.
\begin{figure}
\includegraphics[scale=0.5]{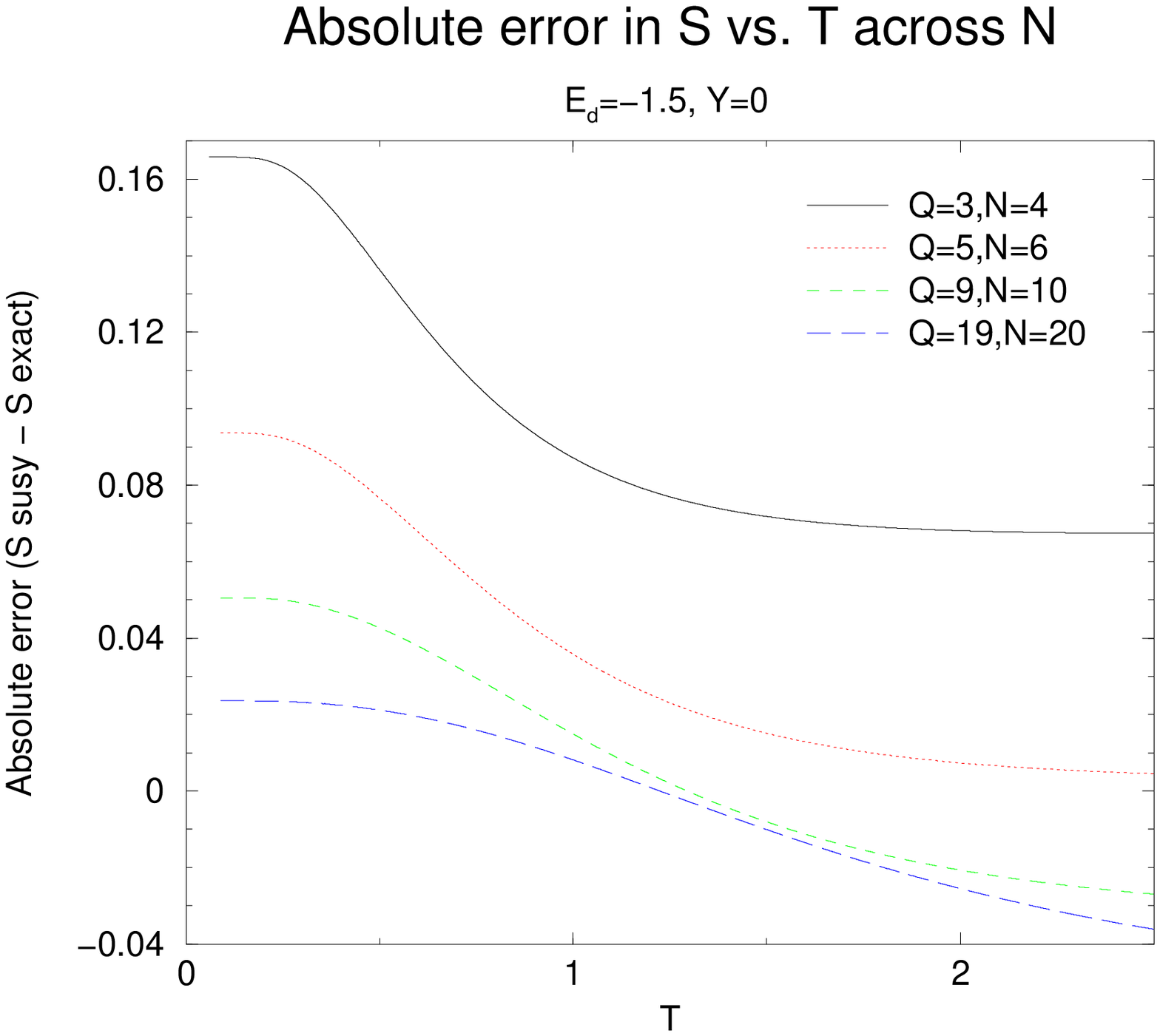}
\includegraphics[scale=0.5]{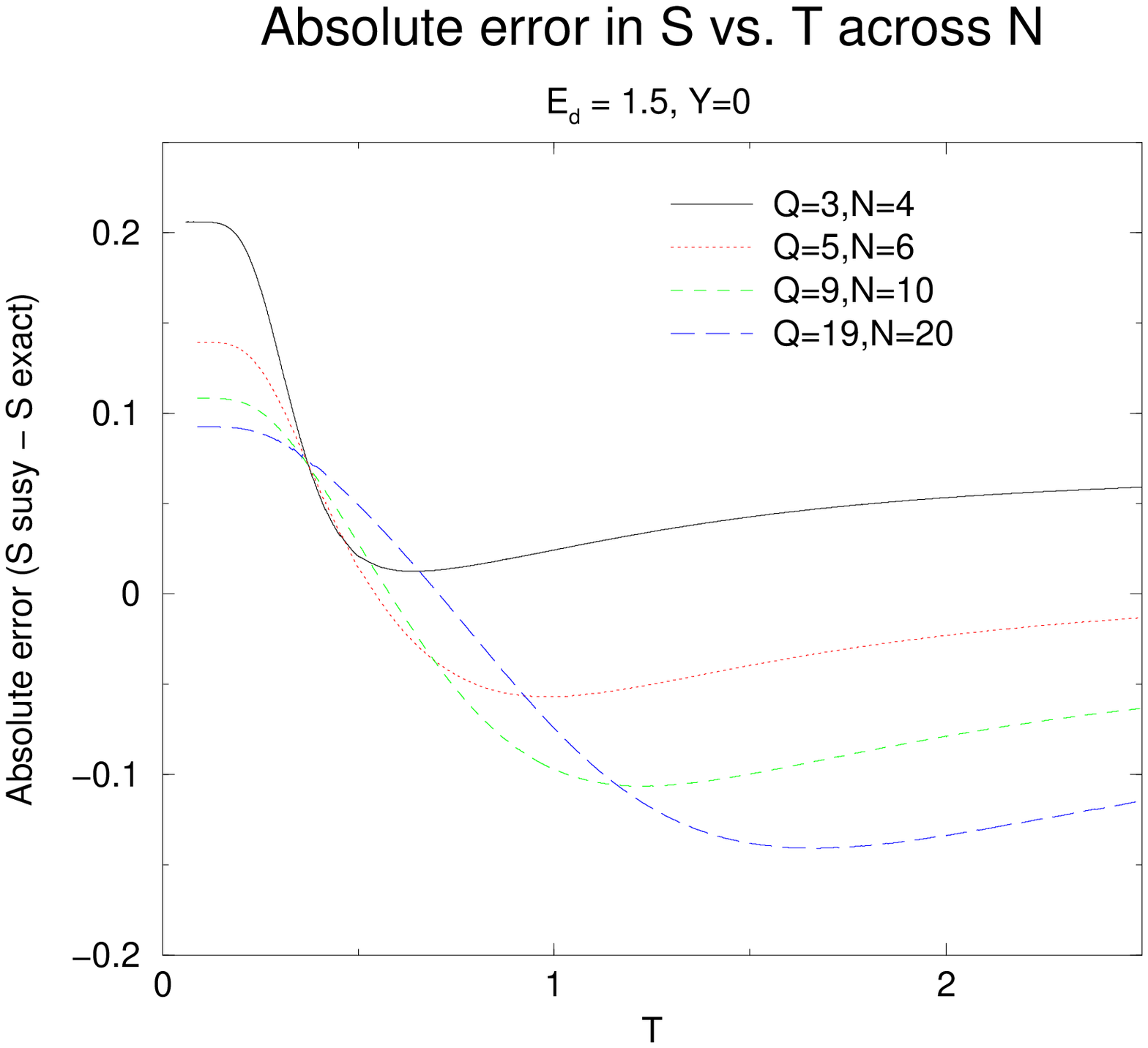}
\caption{\label{figure 10}Absolute error in our entropy approximation as N is increased.(a) $E_d < 0$; (b) $E_d > 0$}
\end{figure}
\begin{figure}
\includegraphics[scale=0.5]{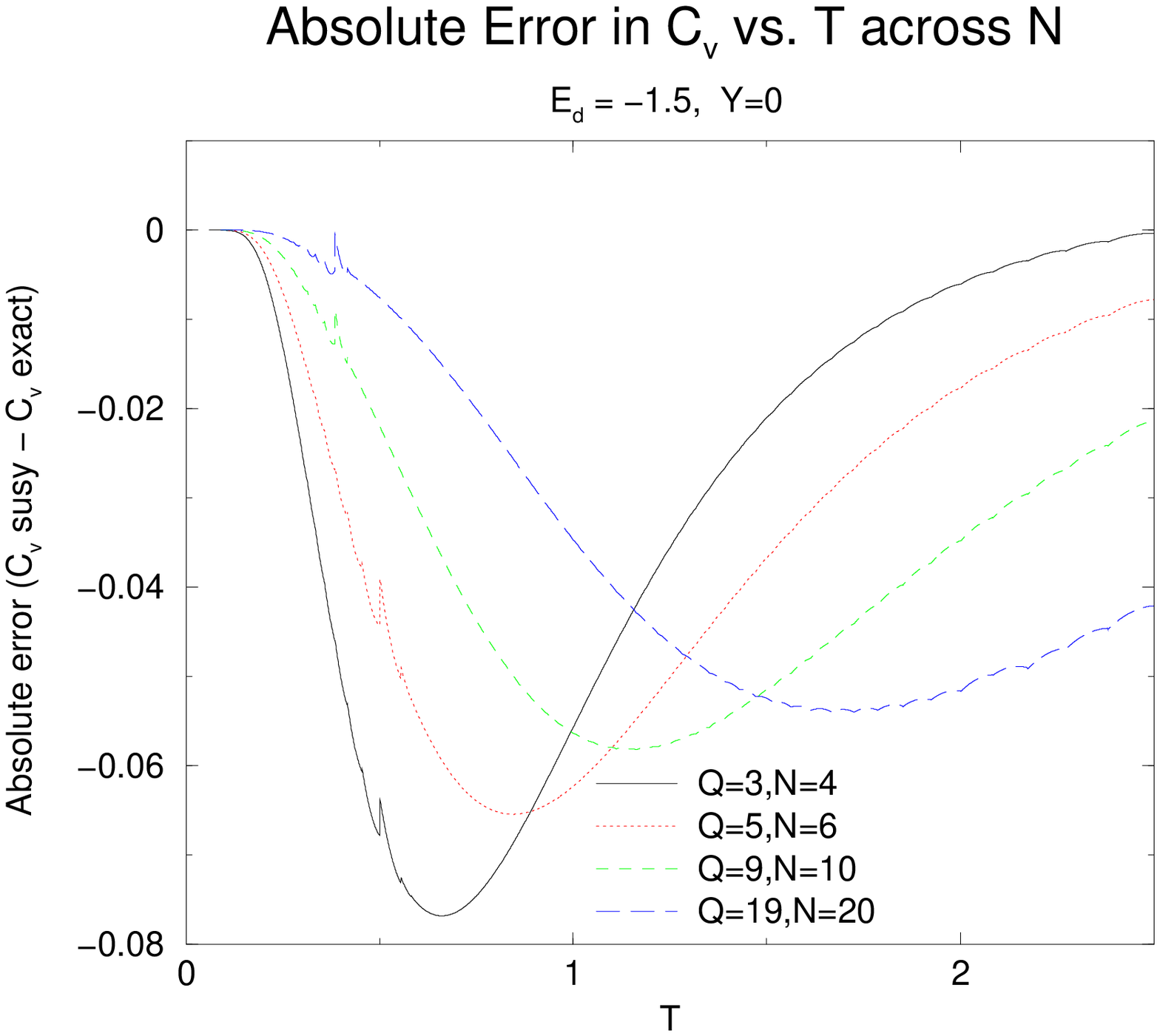}
\includegraphics[scale=0.5]{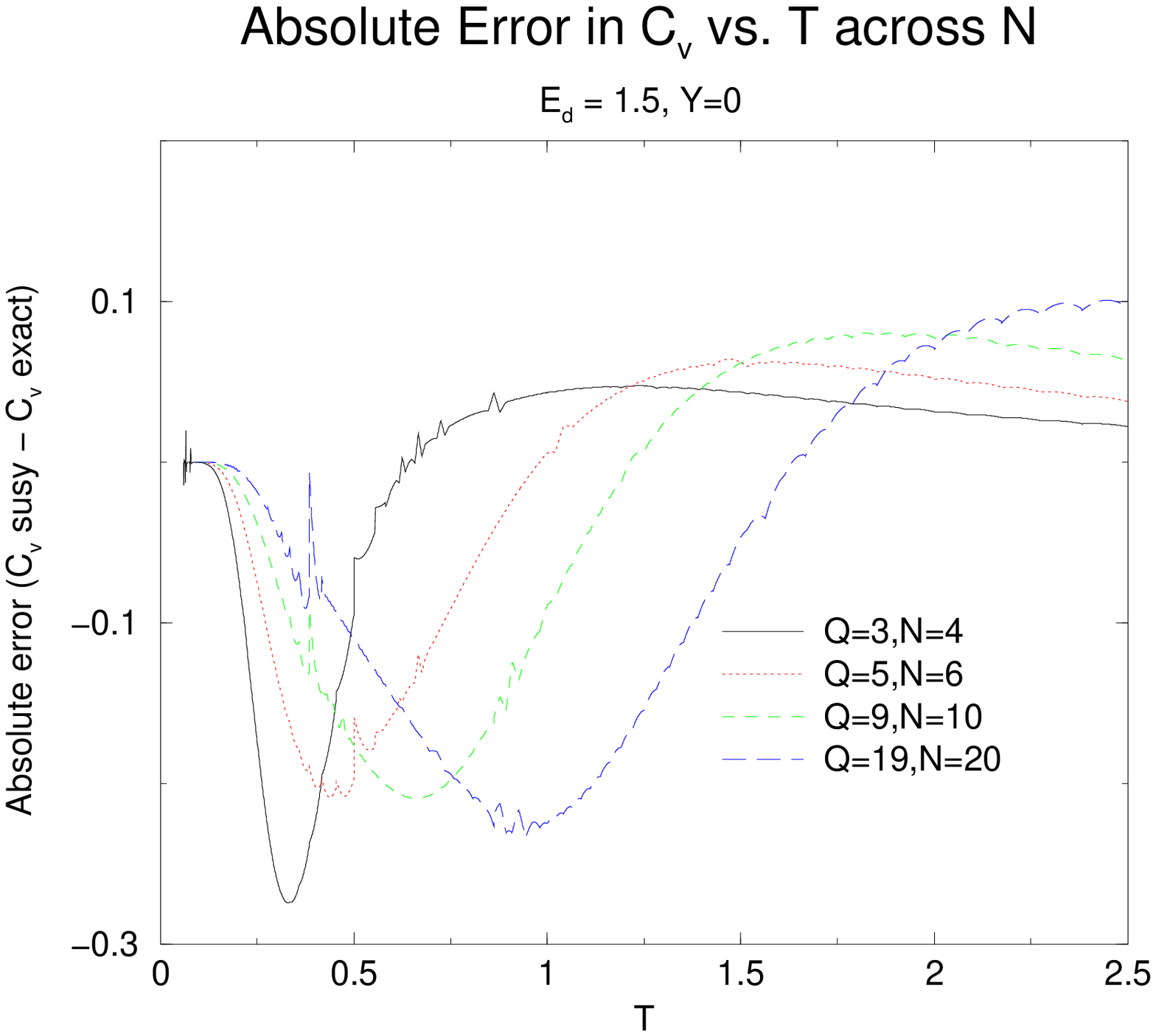}
\caption{\label{figure 11}Absolute error in our heat capacity approximation as N is increased.  The sharp feature is likely a numerical problem where root-finding methods overlap.(a) $E_d < 0$; (b) $E_d > 0$}
\end{figure}

\vskip1pc

{\centering{4. CONCLUSIONS}\\} 

\vskip1pc

We have presented a simple strongly correlated electron model to test three representations of Hubbard operators.  For the limiting case of spin fluctuations, we have demonstrated that mean field theory plus gaussian corrections is capable of capturing the leading order O($\frac{1}{N}$) contributions for our model by comparison with Stirling's approximation expansions of exact results.  Also in this case, by considering a mean field treatment plus gaussian corrections we have shown that a Schwinger boson description produces the most accurate results of the three large N methods.  By introducing charge fluctuations, we have shown that this method of comparison can be extended to calculate heat capacities, and have found that spin states within this model generally are better described by bosonic descriptions (slave fermion) while empty favored by fermionic descriptions (slave boson), in the dangerous limit of N=2.  The lowest energy saddle-point solution arising from the susy representation automatically makes this choice at the mean field level, but encounters a non-physical divergence in its gaussian fluctuations at N=2.  The highest energy saddle-point of the susy representation on the other hand exhibits a qualitative similarity to that seen experimentally by systems near an antiferromagnetic quantum critical point, and it will be interesting to see if adding interactions to this simple model might drive us towards the realization of this type of phase diagram which might be expected to have two distinct bosonic order parameters: charge $\phi_0^2$ which might couple to the hybridization and spin $<b>$ which might allow the realization of a magnetically ordered state.   Furthermore, we have shown that these results can be extended to higher N by the application of simple rules.  We have seen that common to these approaches, the results become exact in the limit of large N, implying that if magnetism and paramagnetism co-exist in the ground-state of a material, at large N one might be justified in using the supersymmetric operators. 

\vskip1pc

{\centering{ACKNOWLEDGEMENTS}\\}

\vskip1pc

Much of the early work on this model was carried out in collaboration with C. P{\'e}pin.  In addition we gratefully acknowledge discussions on this work with C. P{\'e}pin, A. Katanin and O. Parcollet who pointed us towards ref. (22).  This work was supported by National Science Foundation grant DMR 9983156.



\vskip1pc

{\centering{{APPENDIX A: GAUSSIAN APPROXIMATION TO THE ENTROPY OF FREE SPINS}}\\}

\vskip1pc
\setcounter{equation}{0}
\renewcommand{\theequation}{A\arabic{equation}}

The first term of Eqn (7) corresponds to the mean field theory in our approximate large N treatment.
Since we cannot enforce the constraints $Q = Q_0$ and $Y = Y_0$ rigorously at the mean field level, (Y has a term of order $\frac{1}{N}$), we would like to see how inclusion of these terms consistently would affect the saddle point approximation we have made.  The total free energy then is made up of five terms to order $\frac{1}{N}$,
\be
F_{tot} = N F_f + N F_b + F_{\eta} -\lambda Q_0 - \zeta Y_0 +  F_{\delta \lambda_f} + F_{\delta \lambda_b}  = F_{mft} + F_{\frac{1}{N} g. c.}
\ee
where the two terms of order N define the location of the large N saddle point, and the two latter terms are evaluated at the location of the mean field saddle point.  The third term is the expectation value of the fermionic gauge field $\theta$ which we calculate at the level of gaussian fluctuations and then include in the saddle point, while fourth and fifth terms are corrections to the fermionic and bosonic free energies respectively. 
The first and second terms are simply the free fermionic and bosonic free energies:
\be
NF_f = -NT\ln(1 + e^{-\beta \lambda_f}) \hspace{4mm} NF_b = NT\ln(1 - e^{-\beta \lambda_b})
\ee
The third term is the constraint term that we can not strictly handle at the mean field level, $-\frac{\zeta}{Q}<[\theta, \theta^{\dagger}]> = -\frac{2\zeta}{Q}\theta^{\dagger}\theta + \zeta$. To accomplish this, we perform a Hubbard-Stratonovich transformation in the following way:
\be
\frac{2\zeta}{Q_0} (\theta^{\dagger} + \frac{Q_0}{2\zeta} \bar\eta)(\theta + \frac{Q_0}{2\zeta} \eta) -\frac{2\zeta}{Q}\theta^{\dagger}\theta =\frac{Q_0}{2\zeta} \bar\eta \eta + \theta^{\dagger}\eta + \bar \eta \theta
\ee
Since with the introduction of this grassman field we no longer have a quartic term in fields, we can treat these fluctuations at the gaussian level.  This corresponds to calculating
\be
\Pi_{\eta} = \frac{Q_0}{2\zeta} + \includegraphics[scale=1.0]{texteq1.eps} = \frac{Q_0}{2\zeta} + \phi(\omega)
\ee
where
\bea
&\phi(i \omega_n) = -T\sum\limits_{\nu}G_f(\omega + \nu)G_b(\nu) = \oint\frac{dz n(z)}{2 \pi i} \frac{1}{z + i\omega_n + \lambda_f} \frac{1}{z - \lambda_b} \\ & = \frac{n(-i\omega_n + \lambda_f)}{-i\omega_n + 2\zeta} + \frac{n(\lambda_b)}{i \omega_n - 2\zeta} = \frac{f(\lambda_f) + n(\lambda_b)}{\omega - 2 \zeta} = \frac{Q_0}{\omega - 2\zeta}
\eea 
Which means that
\be
\Pi_{\eta} = \frac{Q_0}{2\zeta} + \frac{Q_0}{\omega - 2\zeta} = \frac{Q_0 \omega}{2\zeta (\omega - 2 \zeta)}
\ee
So the contribution to the free energy from these fluctuations is given by
\bea
&F_{\eta} = -\frac{1}{\beta} \sum\limits_{i\omega_n} \ln(\Pi(i\omega_n)) = -\frac{1}{\beta}\sum\limits_{i \omega_n} \ln(\frac{Q_0}{2 \zeta}) + \frac{1}{\beta}\sum\limits_{i \omega_n} \ln(i\omega_n - 2 \zeta) - \frac{1}{\beta}\sum\limits_{i \omega_n} \ln(i\omega_n) \\ &=0 - \oint\frac{dz f(z)}{2 \pi i} \ln(z - 2\zeta) + \oint\frac{dz f(z)}{2 \pi i} \ln(z + 0) \\ &= \frac{1}{\beta}\ln(1 + e^{-2\beta \zeta} - \frac{1}{\beta} \ln(2)) = \frac{1}{\beta}\ln(\frac{1 + e^{-2\beta \zeta}}{2})
\eea
Now, adding back in the additional $\zeta$ from the constraint (as $e^{\ln(\beta \zeta)}$) which was not done for the susy Kondo{\cite{pepin2000,coleman}} model, this becomes $F_{\eta} = T \ln(\frac{e^{\beta \zeta} + e^{-\beta \zeta}}{2})$.  At the mean field level the constraint term reads $-\lambda Q = -\lambda (n_b + n_f - 1)$ which implies an extra contribution of $\lambda$ to the free energy hence this fluctuation term becomes $F_{\eta} = T \ln(\frac{e^{\beta \lambda_f} + e^{\beta \lambda_b}}{2}) = T \ln(\frac{\tilde n_b + \tilde n_f}{2\tilde n_b \tilde n_f})$.  

The corrections coming from varying the Lagrange multiplier about its mean field value and from the new gauge field arise from an expansion of the Lagrangian as $Z = \int{\mathcal D}(f_{\sigma},b_{\sigma}, \delta\lambda_f, \delta\lambda_b, \eta)e^{-\int{d\tau {\mathcal L}_0}}(1 + \delta{\mathcal L} + \frac{1}{2} (\delta {\mathcal L})^2)$ where $L_0$ represents the mean field Lagrangian, and we perform a saddle point evaluation.  For a grassman gauge field, the expansion terminates at second order and the first order terms vanish so we have therefore $e^{(\delta {\mathcal L})^2}$, a grassman gaussian integral.  For the Lagrange multiplier corrections, the first non-zero term is the second order diagram and in the gaussian approximation, we treat this as an exponential correction analogous to the fermionic case.  This being the case, the corrections are as follows.  The diagram of interest for $F_{\delta\lambda_f}$ is :
\be
\includegraphics[scale=1.0]{texteq2.eps}
\ee
This is given by:
\bea
&-T\sum\limits_{i\omega_n \alpha = 1..N}G_f(i\omega_n)G_f(i\omega_n) = \sum\limits_{\alpha} \oint \frac{-dz}{2\pi i} \frac{\partial}{\partial z}f(z) \frac{-1}{z - \lambda_f} \\ &= N( -\beta \tilde n_f(\lambda_f)(1 - \tilde n_f(\lambda_f))  )
\eea
For a real gaussian variable, we have $Z = \int \frac{d(\delta \lambda_f \beta)}{\pi} e^{-N\int_o^{\beta}d\tau \beta \tilde n_f(\lambda_f)(1 - \tilde n_f(\lambda_f))(\delta \lambda_f)^2} = \frac{1}{\pi}\sqrt{\frac{\pi}{2 N \tilde n_f(\lambda_f) (1 - \tilde n_f(\lambda_f))}}$ where the integration is a sum over imaginary time (so it doesn't affect $\tilde n_f$), which totals $\beta$, the inverse temperature. The integration variable is normalized by $\beta$ to make it dimensionless.  The corresponding free energy contribution from these fluctuations is then:
\be
F_{\delta \lambda_f} = -T \ln Z = \frac{T}{2} (\ln(2 \pi N \tilde n_f (1 - \tilde n_f)))
\ee
which is familiar as the gaussian fluctuations about the mean field for Abrikosov fermions.
The diagram of interest for $F_{\delta\lambda_b}$ is:  
\be
\includegraphics[scale=1.0]{texteq3.eps}
\ee
Which is given by:
\bea
&T\sum\limits_{i \nu_n \alpha = 1..N}G_b(i \nu_n) G_b(i \nu_n) = \sum\limits_{\alpha = 1..N}\oint \frac{dz}{2\pi i} n(z) \frac{1}{(z-\lambda_b)^2} \\ &=\sum\limits_{\alpha = 1..N}\oint\frac{-dz}{2\pi i}\frac{\partial n(z)}{\partial z} \frac{-1}{z - \lambda_b} = - N \beta (\tilde n_b(\lambda_b)(1 + \tilde n_b(\lambda_b)))
\eea 
Following the same procedure to find the partition function factor, we get $Z = \frac{1}{\pi}\sqrt{\frac{\pi}{2 N \tilde n_b(\lambda_b) (1 + \tilde n_b(\lambda_b))}}$ so the free energy contribution from these fluctuations is:
\be
F_{\delta \lambda_b} = \frac{T}{2} (\ln(2 \pi N \tilde n_b (1 + \tilde n_b))) 
\ee
which is familiar as the gaussian fluctuations about the mean field for Schwinger bosons.  Enforcing the constraint on Y at the mean field level, couples these two contributions as additional terms arise as $\frac{\partial^2 F_{\eta}}{\partial \lambda_f \partial \lambda_b}$=$\frac{\partial^2 F_{\eta}}{\partial \lambda_b \partial \lambda_f}$=-$\frac{\partial^2 F_{\eta}}{\partial \lambda_b^2}$=-$\frac{\partial^2 F_{\eta}}{\partial \lambda_f^2}$=-$\beta n_{\alpha}(1-n_{\alpha})$ leading to the combined contribution to the gaussian fluctuations as
\be
F_{\delta \lambda_f,\delta \lambda_b} = \frac{T}{2}\ln((2\pi)^2 Det\left[\begin{array}{cc} N \tilde n_f (1 - \tilde n_f) - n_{\alpha}(1-n_{\alpha}) & n_{\alpha}(1-n_{\alpha}) \\ n_{\alpha}(1-n_{\alpha}) & N \tilde n_b (1 + \tilde n_b) -n_{\alpha}(1-n_{\alpha}) \end{array}\right])
\ee
The generalization of this term to the atomic model is straightforward as one incorporates fluctuations in the number of slave partners as well.
\vskip1pc

{\centering{APPENDIX B: Q=2, N=2 ATOMIC WAVEFUNCTIONS}\\}

\vskip1pc
\setcounter{equation}{0}
\renewcommand{\theequation}{B\arabic{equation}}

For Q = 2, N = 2, Y = 1,  we have a totally antisymmetric wavefunction, so to write wavefunctions for these states we start with a simple two-particle anti-symmetric wavefunction and operate with lowering operators ($X_{0 \sigma}$, $\Theta$).  The spin-spin Hubbard operator does not create new states because the initial wavefunctions are totally antisymmetric.  We have with no slaves :
\be
|\hat A> = f^{\dagger}_{\uparrow}f^{\dagger}_{\downarrow}, \hspace{2mm} \Theta |\hat A>
\ee
whose wavefunctions associated are:
\be
f^{\dagger}_{\uparrow}f^{\dagger}_{\downarrow}|0>, \hspace{2mm} b^{\dagger}_{\uparrow}f^{\dagger}_{\downarrow}|0> - b^{\dagger}_{\downarrow}f^{\dagger}_{\uparrow}|0>
\ee
With one vacuum state we have four possibilities:
\be
X_{0 \downarrow} |\hat A>, \hspace{2mm} X_{0 \uparrow} |\hat A>, \hspace{2mm} \Theta X_{0 \downarrow} |\hat A>, \hspace{2mm} \Theta X_{0 \uparrow}|\hat A>
\ee
\be
f^{\dagger}_{\uparrow}\phi^{\dagger}|0>, \hspace{2mm} f^{\dagger}_{\downarrow}\phi^{\dagger}|0>, \hspace{2mm} (b^{\dagger}_{\uparrow}\phi^{\dagger} - \chi^{\dagger}f^{\dagger}_{\uparrow})|0>, \hspace{2mm} (b^{\dagger}_{\downarrow}\phi^{\dagger} - \chi^{\dagger}f^{\dagger}_{\downarrow})|0>
\ee
While with two empty states,
\be
 X_{0 \downarrow} X_{0 \uparrow} |\hat A>, \hspace{2mm}\Theta X_{0 \downarrow} X_{0 \uparrow} |\hat A>
\ee
are the remaining possibilities:
\be
(\phi^{\dagger})^2|0>, \hspace{2mm} \phi^{\dagger}\chi^{\dagger}|0>
\ee


For Q = 2, Y = -1, N = 2, we have a completely symmetric wavefunction, so we naturally start with a bosonic description as the highest weight state and apply boson lowering operators ($X_{0 \sigma}, \Theta^{\dagger}$) and spin exchange operators ($X_{\sigma \sigma}$) to generate the complete set. The fully particle description is comprised of
\be
|B> = b^{\dagger}_{\downarrow}b^{\dagger}_{\downarrow}|0>, \hspace{2mm} X_{\uparrow \downarrow}|B>,  \hspace{2mm} X_{\uparrow \downarrow}X_{\uparrow \downarrow}|B>, \Theta^{\dagger}|B>, \hspace{2mm} \Theta^{\dagger} X_{\uparrow \downarrow}|B>, \hspace{2mm} \Theta^{\dagger}X_{\uparrow \downarrow} X_{\uparrow \downarrow}|B>
\ee
which in terms of fermionic and bosonic operators appears as:
\be
b^{\dagger}_{\downarrow}b^{\dagger}_{\downarrow}|0>, \hspace{2mm} b^{\dagger}_{\uparrow}b^{\dagger}_{\downarrow}|0>, \hspace{2mm}b^{\dagger}_{\uparrow}b^{\dagger}_{\uparrow}|0>,\hspace{2mm} f^{\dagger}_{\downarrow}b^{\dagger}_{\downarrow}|0>,\hspace{2mm} f^{\dagger}_{\uparrow}b^{\dagger}_{\downarrow}|0> + f^{\dagger}_{\downarrow}b^{\dagger}_{\uparrow}|0>,\hspace{2mm}f^{\dagger}_{\uparrow}b^{\dagger}_{\uparrow}|0>
\ee
Only fluctuations to one empty site are possible ($X_{0 \sigma}^2 = 0$)within the constraints for these quantum numbers.  Thus the other possible states are categorized by:
\be
X_{0 \downarrow}|B>, \hspace{2mm} X_{0 \uparrow}|B>, \hspace{2mm} X_{0 \downarrow} \Theta^{\dagger}|B>, \hspace{2mm} X_{0 \uparrow} \Theta^{\dagger}X_{\uparrow \downarrow}X_{\uparrow \downarrow}|B>
\ee
\be 
b^{\dagger}_{\downarrow}\chi^{\dagger}|0>, \hspace{2mm} b^{\dagger}_{\uparrow}\chi^{\dagger}|0>, \hspace{2mm} \phi^{\dagger}b^{\dagger}_{\downarrow}|0> + \chi^{\dagger}f^{\dagger}_{\downarrow}|0>, \hspace{2mm} \phi^{\dagger}b^{\dagger}_{\uparrow}|0> + \chi^{\dagger}f^{\dagger}_{\uparrow}|0>
\ee

\vskip1pc

{\centering{APPENDIX C: Q=3, N=2 ATOMIC WAVEFUNCTIONS}\\}

\vskip1pc
\setcounter{equation}{0}
\renewcommand{\theequation}{C\arabic{equation}}

For the (mixed) atomic model we can treat Q = 3 exactly (without comparison) for all three possible cases (Y = -2,0,2).  Here for comparison's sake we list the wavefunctions of the N = 2 result, so that we can compare with a fairly simple calculation where the corner particle gets to choose its type in a non-trivial way.  Since the degeneracy of the level gives us a value for the partition function, we can easily calculate a free energy and a heat capacity for this atomic model, and compare these results to the algorithm of section II i) d).  For $Q = 3, Y = 0, N = 2$, our basis consists of four states with no slave particles, $8 = 2^3$ distinct states with one slave particle and four states with two slave particles.  For completeness I will list them here, and how they were derived. With no slave particles we have:
\be
|A> = b_{\downarrow}^{\dagger}f^{\dagger}_{\uparrow}f^{\dagger}_{\downarrow}|0>, \hspace{2mm} \Theta|A>, \hspace{2mm} X_{\uparrow \downarrow} |A>, \hspace{2mm} \Theta X_{\uparrow \downarrow}|A>
\ee
which are respectively:
\be
b_{\downarrow}^{\dagger}f^{\dagger}_{\uparrow}f^{\dagger}_{\downarrow}|0>, \hspace{2mm} b^{\dagger}_{\downarrow}(b^{\dagger}_{\uparrow}f^{\dagger}_{\downarrow} |0> - b^{\dagger}_{\downarrow}f^{\dagger}_{\uparrow}|0>), \hspace{2mm} b^{\dagger}_{\uparrow}f^{\dagger}_{\uparrow}f^{\dagger}_{\downarrow}|0>, \hspace{2mm} b^{\dagger}_{\uparrow}(b^{\dagger}_{\uparrow}f^{\dagger}_{\downarrow} |0> - b^{\dagger}_{\downarrow}f^{\dagger}_{\uparrow}|0>).
\ee
With one slave particle:
\bea
&X_{0 \uparrow}& |A>, \hspace{2mm} X_{0 \downarrow}|A>, \hspace{2mm} \Theta X_{0 \uparrow} |A>, \hspace{2mm}\Theta X_{0 \downarrow} |A>, \cr &X_{0 \uparrow}&X_{\uparrow \downarrow}|A>, \hspace{2mm}\Theta X_{0 \uparrow}X_{\uparrow \downarrow}|A>, \hspace{2mm}X_{0 \downarrow}X_{\uparrow \downarrow}|A>, \hspace{2mm}\Theta X_{0 \downarrow}X_{\uparrow \downarrow}|A>
\eea
which are:
\bea
&\phi^{\dagger}b^{\dagger}_{\downarrow}f^{\dagger}_{\downarrow}, \hspace{2mm}\chi^{\dagger}f^{\dagger}_{\uparrow}f^{\dagger}_{\downarrow}|0> - \phi^{\dagger}b^{\dagger}_{\downarrow}f^{\dagger}_{\uparrow}|0>, \hspace{2mm} b^{\dagger}_{\downarrow}(\phi^{\dagger}b^{\dagger}_{\downarrow}|0> - \chi^{\dagger}f^{\dagger}_{\downarrow}|0>), \\ & 2b^{\dagger}_{\downarrow}\chi^{\dagger}f^{\dagger}_{\uparrow}|0> - b^{\dagger}_{\uparrow}(\chi^{\dagger}f^{\dagger}_{\downarrow}|0> + \phi^{\dagger}b^{\dagger}_{\downarrow}|0>), \hspace{2mm} \chi^{\dagger} f^{\dagger}_{\uparrow}f^{\dagger}_{\downarrow}|0> + \phi^{\dagger}b^{\dagger}_{\uparrow}f^{\dagger}_{\downarrow}|0>, \\ &\hspace{2mm} b^{\dagger}_{\downarrow}(\chi^{\dagger}f^{\dagger}_{\uparrow}|0> + \phi^{\dagger}b^{\dagger}_{\uparrow}|0>) -2 b^{\dagger}_{\uparrow}\chi^{\dagger}f^{\dagger}_{\downarrow}|0>, \hspace{2mm} -\phi^{\dagger}b^{\dagger}_{\uparrow}f^{\dagger}_{\uparrow} |0>, \hspace{2mm} b^{\dagger}_{\uparrow}(\chi^{\dagger}f^{\dagger}_{\uparrow}|0> - \phi^{\dagger}b^{\dagger}_{\uparrow}|0>).
\eea
Here we have not written two equivalent eigenvectors.  While with two slave particles:
\be
X_{0 \downarrow}X_{0 \uparrow}|A>, \hspace{2mm} \Theta X_{0 \downarrow}X_{0 \uparrow}|A>, \hspace{2mm} X_{\uparrow \downarrow} X_{0 \downarrow}X_{0 \uparrow}|A>, \hspace{2mm} \Theta X_{\uparrow \downarrow} X_{0 \downarrow}X_{0 \uparrow}|A>
\ee
whose wavefunctions are:
\be
\phi^{\dagger}(\chi^{\dagger}f^{\dagger}_{\downarrow}|0> + \phi^{\dagger}b^{\dagger}_{\downarrow}|0>), \hspace{2mm} b^{\dagger}_{\downarrow}\phi^{\dagger}\chi^{\dagger}|0>, \hspace{2mm} \phi^{\dagger}(\chi^{\dagger}f^{\dagger}_{\uparrow}|0> + \phi^{\dagger}b^{\dagger}_{\uparrow}), \hspace{2mm} b^{\dagger}_{\uparrow}\phi^{\dagger}\chi^{\dagger}|0>
\ee
We can also treat the fully symmetric case Q = 3, Y = -2, N = 2.  As in the case with one less particle, we have spin fluctuations, wavefunction character fluctuations and limited charge fluctuations.  The filled particle levels are described by:
\bea
&|C> = (b^{\dagger}_{\downarrow})^3|0>, \hspace{2mm} X_{\uparrow \downarrow}|C>, \hspace{2mm} X_{\uparrow \downarrow}X_{\uparrow \downarrow}|C>,\hspace{2mm} X_{\uparrow \downarrow}^3|0>, \\ & \Theta^{\dagger}|C>, \hspace{2mm} \Theta^{\dagger}X_{\uparrow \downarrow}|C>, \hspace{2mm}  \Theta^{\dagger}X_{\uparrow \downarrow}X_{\uparrow \downarrow}|C>, \hspace{2mm} \Theta^{\dagger}X_{\uparrow \downarrow}^3|0>
\eea
which are:
\bea
&(b^{\dagger}_{\downarrow})^3|0>, \hspace{2mm} (b^{\dagger}_{\downarrow})^2 b^{\dagger}_{\uparrow}|0>, \hspace{2mm} (b^{\dagger}_{\uparrow})^2 b^{\dagger}_{\downarrow}|0>,(b^{\dagger}_{\uparrow})^3|0> \\ & f^{\dagger}_{\downarrow} (b^{\dagger}_{\downarrow})^2|0>, \hspace{2mm} b^{\dagger}_{\downarrow}(f^{\dagger}_{\uparrow} b^{\dagger}_{\downarrow}|0> + 2 f^{\dagger}_{\downarrow} b^{\dagger}_{\uparrow}|0>), \hspace{2mm}  b^{\dagger}_{\uparrow}(f^{\dagger}_{\downarrow} b^{\dagger}_{\uparrow}|0> + 2 f^{\dagger}_{\uparrow} b^{\dagger}_{\downarrow}|0>), \hspace{2mm} f^{\dagger}_{\uparrow} (b^{\dagger}_{\uparrow})^2|0>
\eea
and with one less:
\bea
&X_{0 \downarrow}|C>, \hspace{2mm} X_{0 \downarrow}X_{\uparrow \downarrow}|C>, \hspace{2mm}X_{0 \downarrow}X_{\uparrow \downarrow}^2|C>,\\ & \hspace{2mm} X_{0 \downarrow}\Theta^{\dagger}|C>, \hspace{2mm}X_{0 \downarrow}\Theta^{\dagger}X_{\uparrow \downarrow}|C>,\hspace{2mm} X_{0 \downarrow}\Theta^{\dagger}X_{\uparrow \downarrow}^2|C>
\eea
\bea
&\chi^{\dagger}(b^{\dagger}_{\downarrow})^2|0>, \hspace{2mm}\chi^{\dagger}b^{\dagger}_{\uparrow}b^{\dagger}_{\downarrow}|0>,\hspace{2mm}\chi^{\dagger}(b^{\dagger}_{\uparrow})^2|0>,\\ & \hspace{2mm} \phi^{\dagger}(b^{\dagger}_{\downarrow})^2|0> + 2\chi^{\dagger} f^{\dagger}_{\downarrow} b^{\dagger}_{\downarrow}|0>, \hspace{2mm}\phi^{\dagger}b^{\dagger}_{\uparrow}b^{\dagger}_{\downarrow}|0> +\chi^{\dagger}( f^{\dagger}_{\uparrow} b^{\dagger}_{\downarrow}|0> +f^{\dagger}_{\downarrow} b^{\dagger}_{\uparrow}|0>), \hspace{2mm} \phi^{\dagger}(b^{\dagger}_{\uparrow})^2|0> + 2\chi^{\dagger} f^{\dagger}_{\uparrow} b^{\dagger}_{\uparrow}|0> 
\eea
And of course we can treat the fully antisymmetric representation (Q=3,Y=2,N=2).  Note that since N = 2 we can no longer have 3 spins!  Here's what we have (totally unoccupied):
\be
|D> = (\phi^{\dagger})^3|0>, \hspace{2mm} \Theta|D>
\ee
which are
\be
(\phi^{\dagger})^3|0>, \hspace{2mm} \chi^{\dagger}(\phi^{\dagger})^2|0>
\ee
one spin
\be
X_{\uparrow 0}|D>, \hspace{2mm}X_{\downarrow 0}|D>, \hspace{2mm}X_{\uparrow 0}\Theta|D>, \hspace{2mm} X_{\downarrow 0}\Theta|D>
\ee
which look like
\be
f^{\dagger}_{\uparrow}(\phi^{\dagger})^2|0>, \hspace{2mm} f^{\dagger}_{\downarrow}(\phi^{\dagger})^2|0>, \hspace{2mm} b^{\dagger}_{\uparrow}(\phi^{\dagger})^2|0> + 2f^{\dagger}_{\uparrow}\chi^{\dagger}\phi^{\dagger}|0>, \hspace{2mm} b^{\dagger}_{\downarrow}(\phi^{\dagger})^2|0> + 2f^{\dagger}_{\downarrow}\chi^{\dagger}\phi^{\dagger}|0>
\ee
two spins:
\be
X_{\uparrow 0}X_{\downarrow 0}|D>, \hspace{2mm} X_{\uparrow 0}X_{\downarrow 0}\Theta|D>
\ee
and these are:
\be
f^{\dagger}_{\uparrow}f^{\dagger}_{\downarrow}\phi^{\dagger}|0>, \hspace{2mm}\phi^{\dagger}(f^{\dagger}_{\downarrow} b^{\dagger}_{\uparrow}|0> -f^{\dagger}_{\uparrow} b^{\dagger}_{\downarrow}|0>) + \chi^{\dagger}f^{\dagger}_{\downarrow}f^{\dagger}_{\uparrow}|0>
\ee 

{\centering{APPENDIX D: MEAN FIELD OCCUPATION}\\}

\vskip1pc
\setcounter{equation}{0}
\renewcommand{\theequation}{D\arabic{equation}}

To gain an appreciation of how the character of the spin system changes as a function of temperature, in Fig. 12 we show the mean field results of the slave fermion approximation for Q = 1, N=2.  We see (Fig. 12 (a)) that in this approximation when $E_d > 0$ the Hubbard operators describe a charged hole which obeys grassman statistics.  As the temperature is raised the bosonic description of the spin begins to carry more weight.  When $E_d<0$, the ground state at T=0 is a pure spin state which acquires a bosonic description (Schwinger bosons) (Fig. 12 (b)).
\begin{figure}
\includegraphics[scale=0.4]{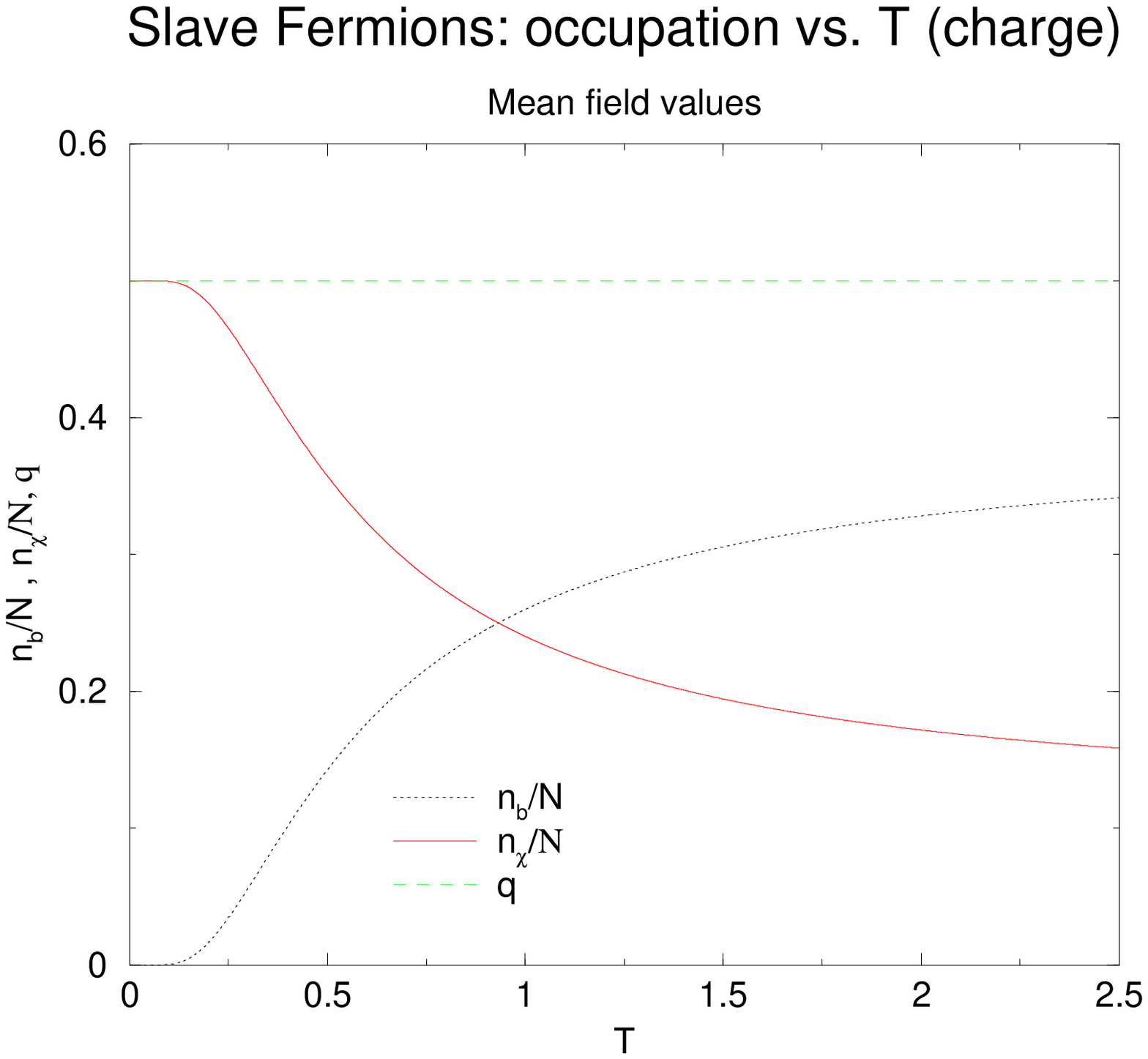}
\includegraphics[scale=0.4]{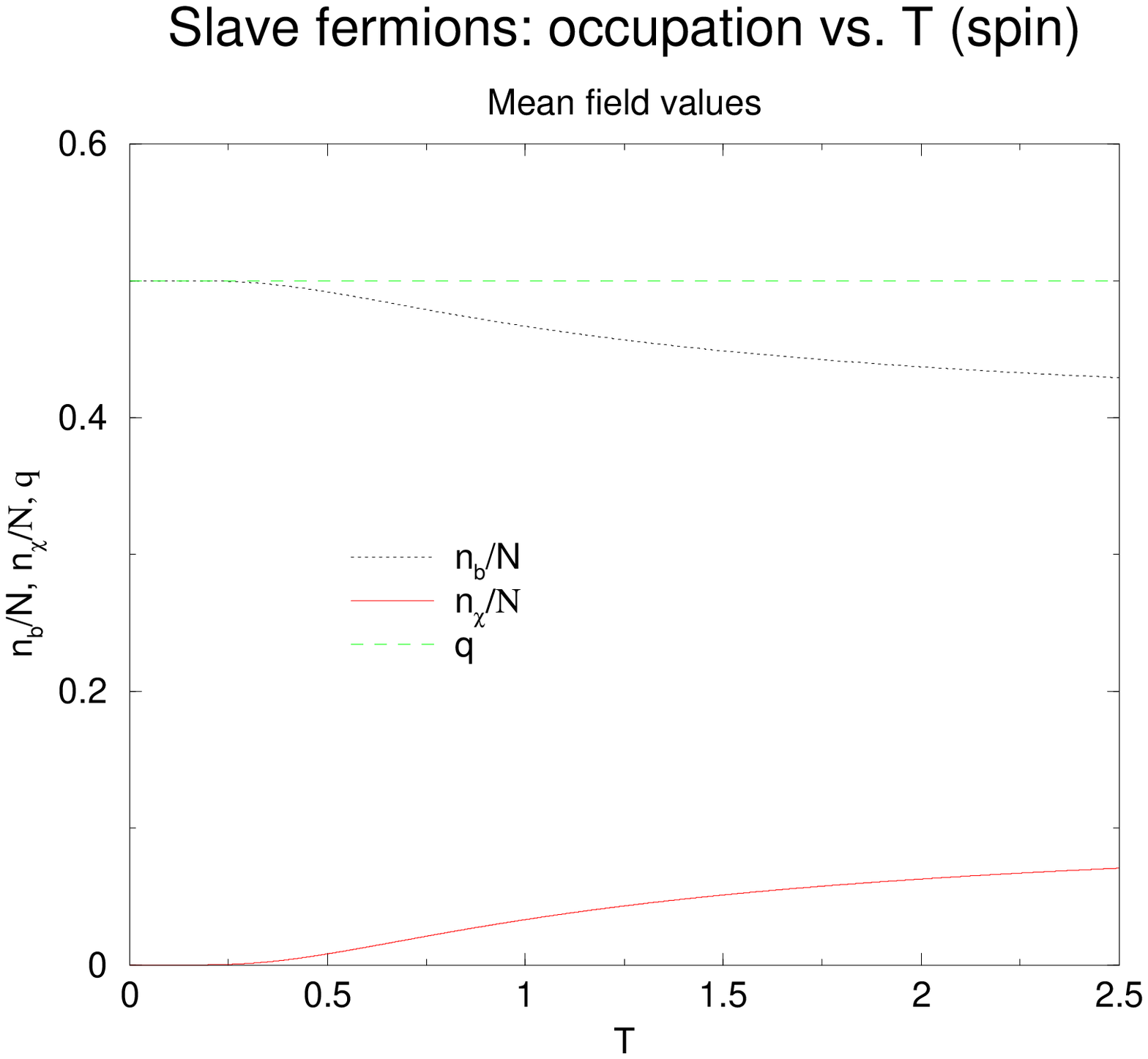}
\caption{\label{figure 12}Slave fermions: (a) $E_d = 1.5$ (empty); (b) $E_d = -1.5$  }
\end{figure}
In Fig. 13 we show the mean field results of the slave boson approximation for Q=1, N=2.  In Fig. 13 (a) we see that for $E_d > 0$ the zero temperature ground state of the system is a positively charged bosonic hole.  As the temperature is increased some fermionic spinorial character of the electron returns.  When $E_d < 0$ the system enters a fermionic ground state at low temperatures which is just the free spin case treated in section 2 (Abrikosov fermions). (see Fig. 13 (b)).

\begin{figure}
\includegraphics[scale=0.4]{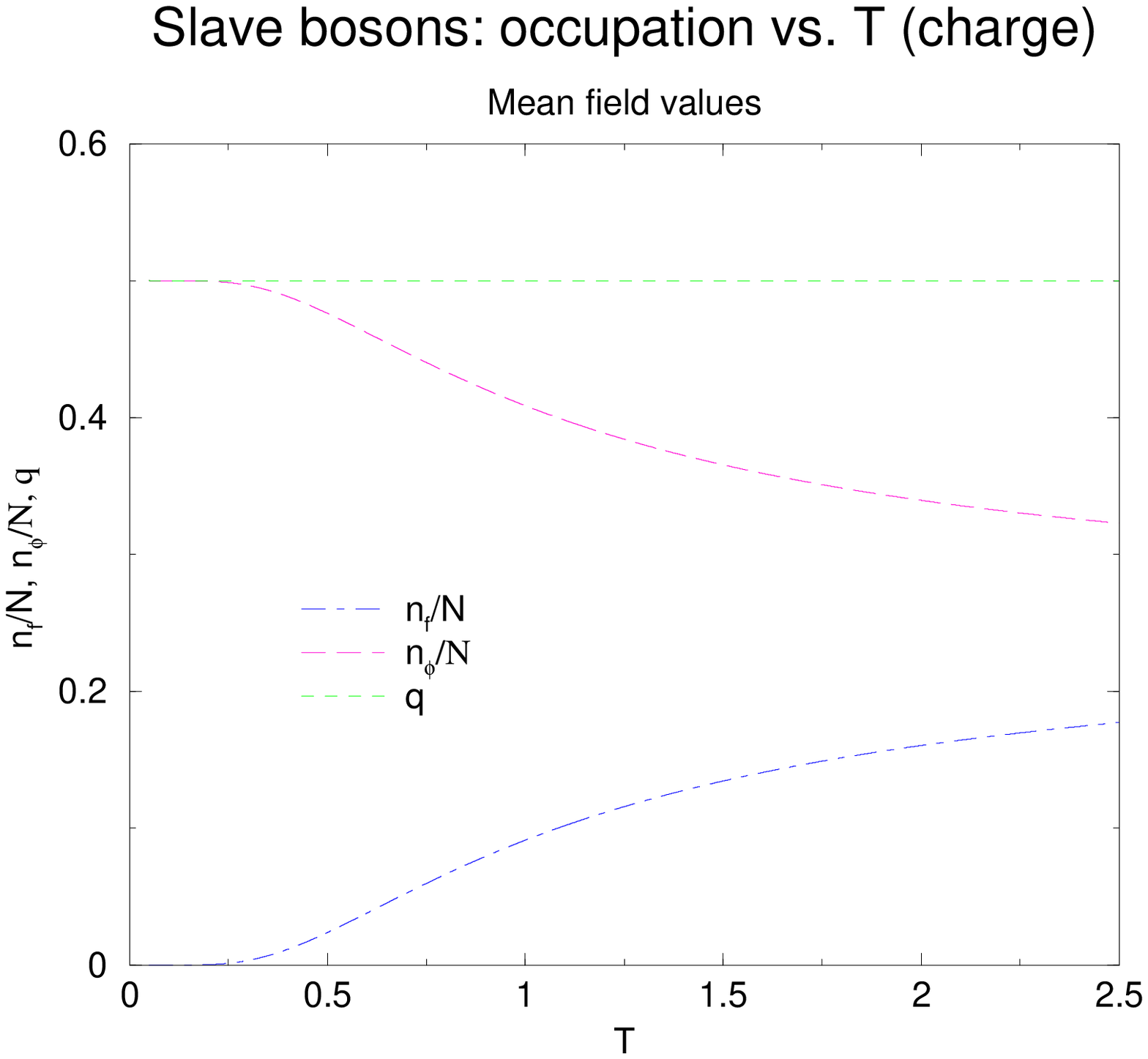}
\includegraphics[scale=0.4]{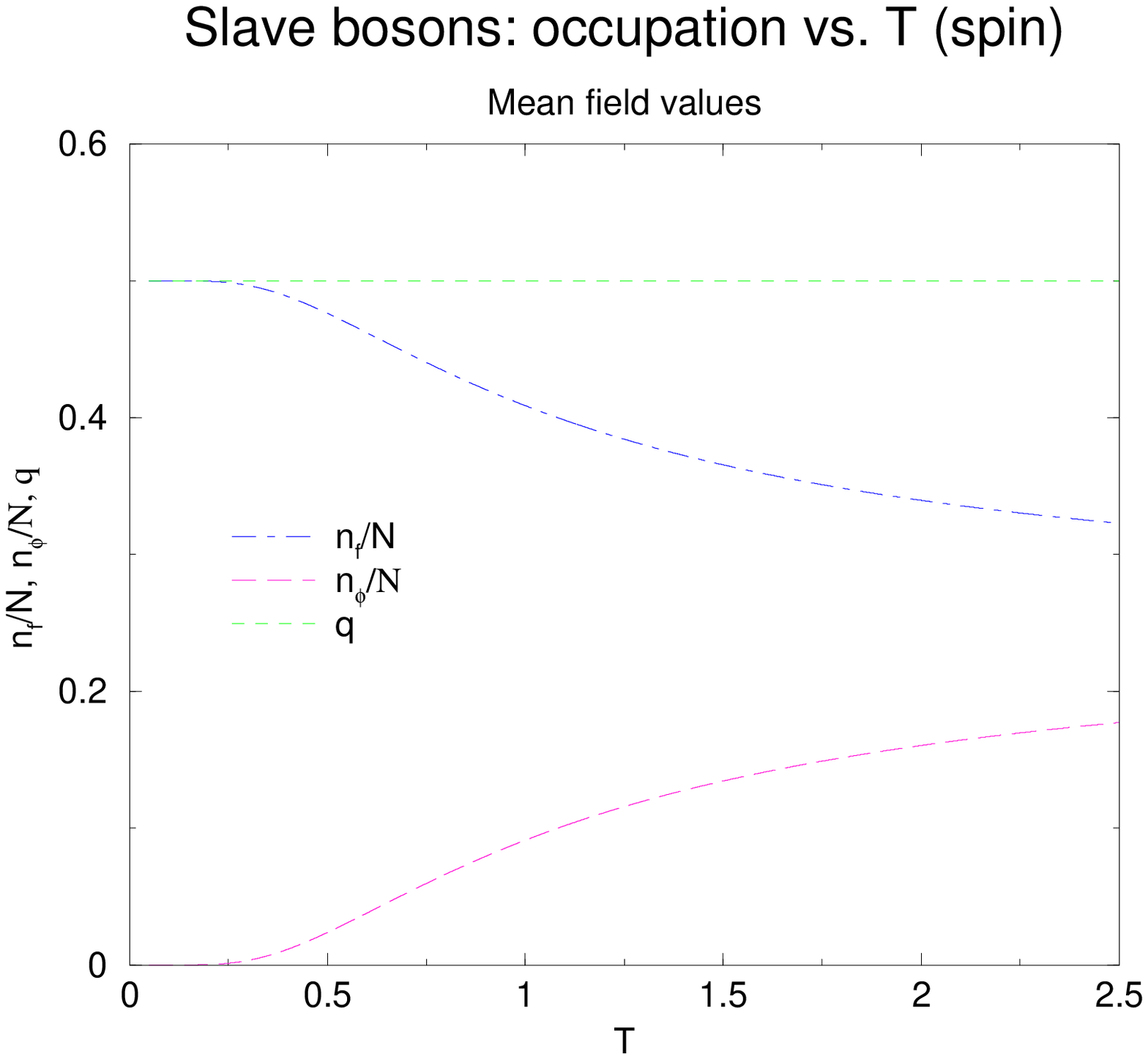}
\caption{\label{figure 13}Slave bosons: (a) $E_d = 1.5$; (b) $E_d = -1.5$ }
\end{figure}

For the case of physical interest (Q=1,N=2,Y=0), we saw that the supersymmetric description has three possible physical analytic solutions (see Table III).  When $n_{\alpha} = 0$ one recovers the slave fermion mean field results (Fig. 12), when $n_{\alpha} = 1$ one recovers the slave boson results (Fig. 13), and a mixed solution exists in the small region $E_d = 0$ to $E_d = T\ln(2)$ as shown in the inset to Fig. 14.  If one follows the Y=0 series by increasing N while holding $\tilde h = \frac{h}{N} = \frac{1}{2}$ constant (see Fig.14), one sees that one generically has a mixed solution -- neither slave bosons nor slave fermions can describe the physics of systems exhibiting this symmetry.  As approximated methods become asymptotically exact as N $\rightarrow \infty$, if the phase diagram of Fig. 6 (b) were to become the physically realized state, one might be able to justify a large N expansion to treat the physics of the mixed region using susy Hubbard operators at Y=0.

\begin{figure}
\includegraphics[scale=0.5]{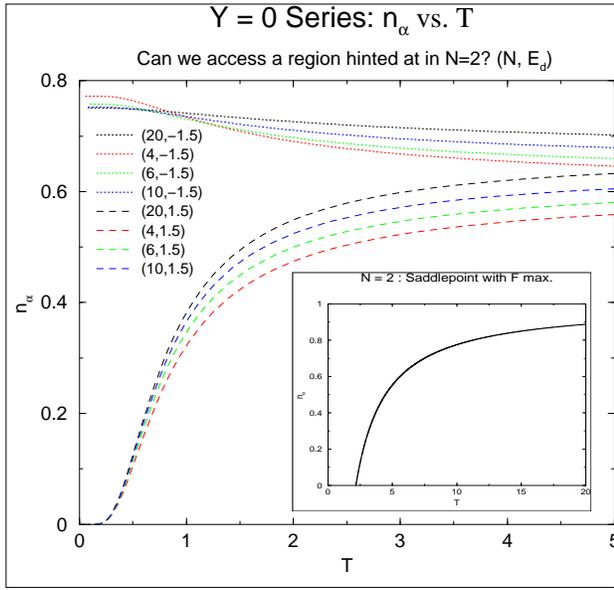}
\caption{\label{figure 14}The Y = 0 series.  In the free spin limit ($E_d/T << 0$), the corner box favors a fermionic description of the spin (slave bosons $n_{\alpha} = 1$)  while in the free charge limit ($E_d/T >> 0$), the corner box favors a fermionic description of the charge (slave fermions $n_{\alpha} = 0$ ).  At high temperatures a mixed solution exists slightly favoring slave bosons.  Both the heat capacity and entropy approximations for N$>$2 produce physical results and may be seen as an extension of the high temperature N=2 mixed solution (see inset) which for $E_d > 0$ and $T>\frac{E_d}{\ln(2)}$ tunes from slave fermion to slave boson as temperature increases.  Study of the physics of this series in less trivial models may reproduce some of the interesting physics seen close to a quantum critical point.}
\end{figure}

\end{document}